\documentclass[10pt, two column, twoside]{IEEEtran}
\usepackage{lipsum}
\usepackage{cuted}
\usepackage{amsthm}
\usepackage{amsmath}
\usepackage{algorithmic}
\usepackage{algorithm}
\usepackage{mathrsfs}
\usepackage{graphicx}
\usepackage{subfigure}
\usepackage{cite}
\usepackage{bm,epsfig,amsthm,url}
\usepackage{indentfirst}
\usepackage{amssymb}
\usepackage{epstopdf}

\usepackage{xcolor}
\usepackage{mathtools}
\usepackage{hyperref}
\usepackage{verbatim}
\usepackage{enumerate} 

\usepackage{makecell}
\theoremstyle{definition}

\usepackage[final]{changes} 
\definecolor{MyColor}{RGB}{0, 0 , 255}
\definecolor{MyColor2}{RGB}{0, 0 , 0}
\definecolor{MyColor3}{RGB}{0, 255 , 0}

\begin{document}

\title{Multi-IRS Enhanced Wireless Coverage: Deployment Optimization Based on Large-Scale Channel Knowledge}
\author{Min~Fu,~\IEEEmembership{Member,~IEEE}, Lipeng~Zhu,~\IEEEmembership{Member,~IEEE}, and~Rui~Zhang,~\IEEEmembership{Fellow,~IEEE}
	\thanks{M. Fu and L.~Zhu are with the Department of Electrical and Computer Engineering, National University of Singapore, Singapore 117583 (e-mails:  fumin@nus.edu.sg and zhulp@nus.edu.sg).}
			\thanks{R. Zhang is with School of Science and Engineering, Shenzhen Research Institute of Big Data, The Chinese University of Hong Kong, Shenzhen, Guangdong 518172, China (e-mail: rzhang@cuhk.edu.cn). He is also with the Department of Electrical and Computer Engineering, National University of Singapore, Singapore 117583 (e-mail: elezhang@nus.edu.sg).
}
}

\maketitle

\setlength\abovedisplayskip{2pt}
\setlength\belowdisplayskip{2pt}
\setlength\abovedisplayshortskip{2pt}
\setlength\belowdisplayshortskip{2pt}
\setlength\arraycolsep{2pt}

\begin{abstract}

In this paper, we study the intelligent reflecting surface (IRS) deployment problem where a number of IRSs are optimally placed in a target area to improve its signal coverage with the serving base station (BS).
To achieve this, we assume that there is a given set of candidate sites in the target area for deploying IRSs and divide the area into multiple grids of identical size. Then, we derive the average channel power gains from the BS to IRS in each candidate site and from this IRS to any grid in the target area in terms of IRS deployment parameters, including its size, position, height, and orientation. 
Thus, we are able to approximate the average cascaded channel power gain from the BS to each grid via any IRS, assuming an effective IRS reflection gain based on the large-scale channel knowledge only.
Next, we formulate a multi-IRS deployment optimization problem to minimize the total deployment cost by selecting a subset of candidate sites for deploying IRSs and jointly optimizing their heights, orientations, and numbers of reflecting elements while satisfying a given coverage rate performance requirement over
 all grids in the target area.
To solve this challenging combinatorial optimization problem, we first reformulate it as an integer linear programming problem and solve it optimally using the branch-and-bound (BB) algorithm. 
In addition, we propose an efficient successive refinement algorithm to further reduce computational complexity.
Simulation results demonstrate that the proposed lower-complexity successive refinement algorithm achieves near-optimal performance but with significantly reduced running time compared to the proposed optimal BB algorithm, as well as superior performance-cost trade-off than other baseline IRS deployment strategies.

\end{abstract}
\begin{IEEEkeywords}
Intelligent reflecting surface (IRS), IRS deployment, network coverage, large-scale channel knowledge, integer linear programming.
\end{IEEEkeywords}

\section{Introduction}

 In recent years, significant advancements have been made in the development, implementation, and application of digitally controlled metasurfaces. 
 These metasurfaces enable dynamic manipulation of  electromagnetic waves through tunable passive reflection. Among them, intelligent reflecting surface (IRS)  has emerged as a key technology for wireless communications. Specifically, IRS consists of a large number of reflecting elements, each of which can be dynamically tuned to alter the amplitude and phase of its reflected signal via a smart controller \cite{Wu2021Tutorial}. By adaptively adjusting the amplitudes and/or phase shifts of massive passive reflecting elements, IRS can effectively reshape wireless channels to enhance communication performance in a cost-efficient manner. Unlike traditional base stations (BS) or relays comprising active components, IRS reflection does not require dedicated transmit/receive radio frequency (RF) chains, thus significantly reducing the hardware cost and energy consumption. This makes IRS an appealing solution for dense deployment in wireless networks for improving communication or sensing performance \cite{Mei2022Multireflection}. 
 The promising advantages of IRS have spurred extensive research on its various implementation aspects, such as reflection optimization, channel estimation, and deployment design (see \cite{Wu2021Tutorial, Mei2022Multireflection, Wu2024} and the references therein).
 
  \begin{figure}[t]
 	\centering		
 	\includegraphics[scale=0.3]{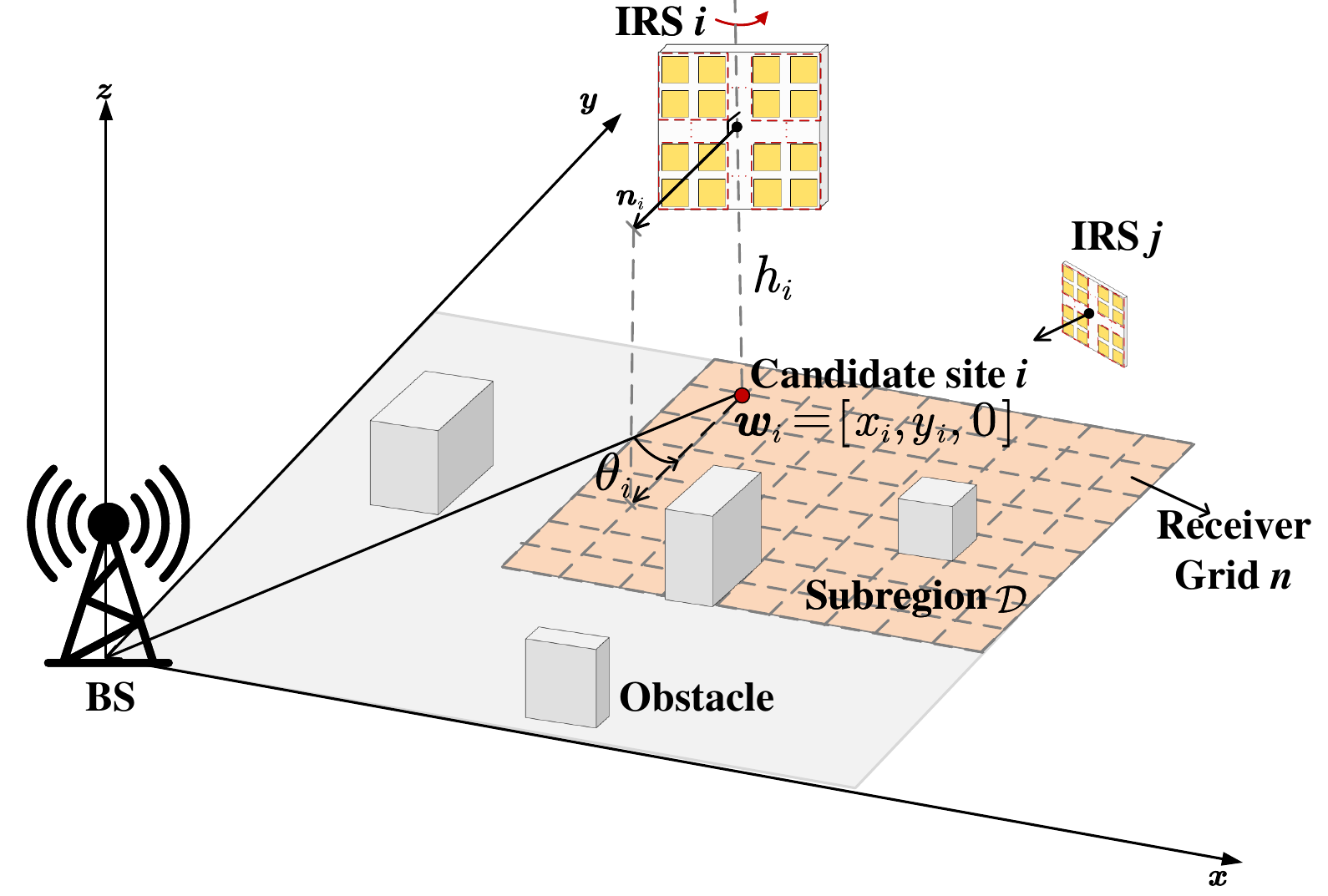}	
 	\vspace{-3mm}		
 	\caption{Illustration of IRS deployment in a 3D environment.} \label{Fig:3Dlayout}	
 	\vspace{-6mm}
 \end{figure}

In particular, to fully achieve the benefits of IRS, the deployment of IRS in wireless networks should be optimally designed by considering its practical half-space reflection  and the specific  propagation environment
 \cite{Wu2021Tutorial, Mei2022Multireflection, Wu2024}.
 To achieve this end, there have been substantial prior works addressing the deployment optimization for IRS in different system setups.
 For example, in the case of passive IRS (PIRS), the authors in \cite{Mu2021SingleReflection} studied a PIRS location optimization problem to maximize the weighted sum rate in multi-user communications under three different multiple access schemes.
  Moreover,   the authors in \cite{Lu2021SingleReflect} and \cite{TANG2021Jamming} considered deployment optimization problems for an unmanned aerial vehicle (UAV)-mounted PIRS.
 In addition to the PIRS's location optimization, the authors in \cite{Zeng2021Orientation, Cheng2022Aerial, MO2024Reconfigurable} further optimized its rotation and showed their joint optimization gain  in terms of performance enhancement.
 Furthermore, the authors in \cite{Hashida2020PIRS,  LIU2021RIS, Huang2022multiplePIRS, Huang2022cost,Bai2022SingleReflection} delved into the PIRS deployment design in a given area and aimed to optimize the locations and/or number of PIRSs deployed.
In addition, the authors in \cite{You2021AIRS} and \cite{kang2023active} explored the deployment problems considering a new IRS architecture, i.e., active IRS (AIRS), which can amplify the reflected signal besides phase shifting. 
Furthermore, the works \cite{FU2023MultiActive1} and \cite{Fu2022active} considered a given multi-IRS reflection path formed by one AIRS and multiple PIRSs, and optimized the AIRS's location over this path jointly with all IRSs' beamforming to maximize the received signal-to-noise ratio (SNR) at users.

 The above studies primarily focused on single-IRS deployment optimization.
To overcome this limitation, recent works have investigated the more general multi-IRS deployment problem to enhance  the wireless coverage performance in complex environments with dense and scattered obstacles.
For example, in \cite{KISHK2021Exploiting}, the authors analyzed the impact of large-scale IRS deployment on the coverage probability in cellular networks via stochastic geometry by assuming randomly located blockages.
In \cite{Efrem2023IRSdeploy}, the authors jointly optimized the locations and sizes of  IRSs to minimize outage probability in multi-IRS aided two-way full-duplex communication systems. Additionally, \cite{Mei2023IRSdeploy} modeled the multi-IRS location optimization problem using graph theory to enhance wireless coverage, aiming to minimize the number of signal reflections per multi-IRS link. Furthermore, a more general graph-based system model was proposed in \cite{fu2024multi1} for the optimization of multi-passive/active-IRS deployment, where the impact of IRSs' locations and sizes on the SNR performance at any user location within a given area was characterized.

 However, the existing studies on multi-IRS deployment\cite{KISHK2021Exploiting,Efrem2023IRSdeploy,Mei2023IRSdeploy,fu2024multi1} assumed  line-of-sight (LoS) links as well as fixed orientations and heights for IRSs deployed. 
Since IRS can only reflect signals to/from its front half-space, its orientation and height should be properly designed to achieve optimum coverage performance.
Thus, it still remains unknown how to jointly optimize the positions, sizes, orientations, and heights of multiple IRSs under a general multi-path propagation environment, which thus motivates our current work. 
In this paper, we study a multi-IRS deployment problem where a number of IRSs are optimally placed in a target area to improve its signal coverage with the serving BS.
 The main contributions of this paper are summarized as follows:
 \begin{itemize}
 	\item %
 	 To facilitate IRS deployment, we assume that there are a given set of candidate sites in the target area for IRS deployment and divide this area into multiple grids of identical size, with each grid serving as a basic unit in our analysis.
 Furthermore, we adopt the average channel power gain over all locations within each grid as the performance measurement to eliminate the effect of multi-path small-scale fading in it.
 Building upon this model, we derive the average channel power gain for the link from the BS to IRS in each candidate site and from this IRS to any grid in the target area, in terms of IRS deployment parameters, including its position, size, height, and orientation.
 Thus, we can approximate the average cascaded  channel power gain from the BS to each grid via any IRS by assuming an effective IRS reflection gain based on the large-scale channel knowledge only.

 	\item Next, we formulate a multi-IRS deployment optimization problem to select an optimal subset of candidate sites for deploying IRSs, which is jointly optimized with their heights, orientations, and numbers of reflecting elements, such that the total deployment cost is minimized while ensuring a given coverage rate performance over all grids in the target area.
 	To solve the formulated combinatorial optimization problem, we reformulate it as an integer linear programming (ILP) problem and then solve it optimally using the branch-and-bound (BB) algorithm.
 	
 	\item In addition, we propose a lower-complexity successive refinement algorithm to further enhance computational efficiency.
 	Its basic idea is to iteratively solve a sequence of subproblems derived from the original optimization problem by successively refining and reducing a subset of candidate sites, until no further reduction in deployment cost can be achieved.
 	Furthermore, to solve each subproblem, we propose a sequential IRS deployment algorithm with a polynomial complexity, which sequentially selects  the highest-priority candidate site for deploying IRS and determines its deployed tile number, height, and orientation in closed form, until the target coverage rate is achieved.
 	
 	 \item Simulation results show that our proposed lower-complexity algorithm achieves near-optimal performance but with significantly reduced running time compared to the proposed optimal BB algorithm. Additionally, the proposed joint deployment optimization strategy with the optimal or suboptimal algorithm outperforms other baseline IRS deployment strategies in terms of the overall deployment cost.

 \end{itemize}

 The remainder of this paper is organized as follows.
 Section \ref{Sec_Model} describes the system model.
 Section \ref{Sec_formulation} characterizes the IRS deployment cost and coverage rate performance in the considered system and presents the problem formulation for multi-IRS deployment optimization.
 Section \ref{Sec:Solution} presents the optimal solution and an efficient suboptimal solution to the formulated problem.
 Section \ref{Sec_Results} presents numerical results to evaluate the effectiveness of our proposed optimal and suboptimal algorithms.
 Section \ref{Sec_Conclusion} concludes this paper.
 
\textit{Notations}:
$\tbinom{n}{k}$ denotes the number of combinations to choose $k$ elements from a set of $n$ elements.
$\mathbb{N}^+$ denotes the set of positive integers.
$\jmath$ denotes the imaginary unit of a complex number.
$x\sim \mathcal{CN}(0, \sigma^2)$ 
represents a circularly symmetric
complex Gaussian random variable $x$ with zero mean and variance $\sigma^2$.
$\mathbb{E}(\cdot)$ denotes the statistical expectation.
$(\cdot)^{\sf H}$ and $(\cdot)^{\sf T}$ denote the conjugate transpose and transpose, respectively.
For a complex-valued vector $\bm x$, $\|\bm x\|$ denotes its Euclidean norm. 
 $\text{diag}(\bm z)$ denotes a diagonal matrix with each diagonal entry being the corresponding element in vector $\bm z$.
For a set $\cal{S}$, $|\cal{S}|$ denotes its cardinality. $\emptyset$ denotes an empty set.
For two sets $\cal{S}$ and $\cal{S}'$, $\cal{S}\cap \cal{S}'$ denotes their intersection, $\cal{S}\cup \cal{S}'$ denotes their union, and $\cal{S}\backslash \cal{S}'$ is the set of elements that belong to $\cal{S}$ but are not in $\cal{S}'$.

\section{System Model}\label{Sec_Model}

As depicted in Fig. \ref{Fig:3Dlayout}, we study an IRS-enhanced wireless system in a given sub-region of interest, denoted as ${\cal D}$. 
Specifically, a BS is located at the center of the cell to serve the users in the sub-region, which is assumed to be far from the BS and thus suffer weak signal coverage from it\footnote{In this paper, we assume that the BS's antenna directional gain towards the sub-region of interest is fixed and thus the average channel power gain between them is a priori known.}.
To improve signal coverage without allocating more transmit power or antennas at the BS, a number of IRSs need to be deployed in ${\cal D}$ to enhance the average channel power between the BS and all possible user locations within it.
To facilitate the deployment design, we assume that $I_0$ candidate sites have been identified in ${\cal D}$\footnote{These candidate sites are preselected which typically have favorable large-scale channel power gains between them and the BS  as well as practical suitability to install IRS  of a sufficiently large size.}, each of which can potentially accommodate an IRS with a flexible size.
In addition, the height and orientation of the IRS at each candidate site are assumed to be generally adjustable through a mechanical control for a more flexible deployment design.
We assume that deploying IRSs with their maximum allowable sizes at all candidate sites can achieve the desired coverage performance from the BS to ${\cal D}$.
However, such dense IRS deployment is cost-prohibitive in practice. Therefore, we aim to select only a (small) subset of these candidate sites for IRS deployment to minimize the deployment cost while satisfying the given signal coverage requirement in ${\cal D}$.

\subsection{IRS Deployment Configuration}
In this subsection, we model the IRS deployment configuration including the IRS's site (i.e., position), size, height, and orientation. As illustrated in Fig. \ref{Fig:3Dlayout}, we establish a global three-dimensional (3D) coordinate system, with the $x$-$O$-$y$ plane representing the plane in which the region ${\cal D}$ is located, and the origin $O$ set at the BS's location.
Let ${\cal I}_0 \triangleq \{1,\ldots,I_0\}$ represent the set of candidate IRS sites in ${\cal D}$, and  ${\bm w}_i \triangleq [x_{i}, y_{i}, 0]^{\sf T} \in \mathbb{R}^3$ denote the coordinates of candidate IRS site $i$, where $i \in {\cal I}_0$. 
Furthermore, we denote by ${\cal I}$ the subset of candidate sites selected to deploy IRSs, i.e., ${\cal I} \subseteq {\cal I}_0$. For convenience, we refer to the IRS deployed at site $i$, $i \in {\cal I}$, as IRS $i$.

In this paper, we consider that the IRS at any candidate site can adjust its altitude, as depicted in Fig. \ref{Fig:3Dlayout}. Specifically, the height of IRS $i$, $i \in {\cal I}$, over the $x$-$O$-$y$ plane, denoted as $h_{i}$, is subject to the following constraint:
\begin{eqnarray} \label{Cons:height}
	h_{i} \in {\cal H}_i, \
\end{eqnarray}
where ${\cal H}_i\triangleq \{\tilde{h}_{i,1},\ldots, \tilde{h}_{i,P_i} \}$ denotes the set of $P_i$ possible discrete heights for IRS $i$, $i \in {\cal I}$.

Furthermore, since the IRS can only reflect signals to/from its front half-space, the orientation of each IRS is another important parameter impacting its signal reflection and influencing the coverage performance.
We assume that the IRS plane is perpendicular to the ground and thus its orientation can only be adjusted in the horizontal plane, as shown in Fig. \ref{Fig:3Dlayout}. Accordingly, let ${\bm n}_i \triangleq [{\bm n}_{i}[x],{\bm n}_{i}[y],0]^{\sf T}\in \mathbb{R}^3$ denote the normal vector of IRS $i$ in the 3D coordinate system, pointing outward from the IRS.
Moreover, let $\theta_i$ represent the azimuth rotation angle with respect to (w.r.t.) $\bm w_i$, as shown in Fig. \ref{Fig:3Dlayout}. In this way, we express ${\bm n}_i$ as a function of $\theta_i$, i.e.,
\begin{eqnarray} \label{eq:normvector}
	{\bm n}_i(\theta_i) = \begin{bmatrix}
		\cos(\theta_i) & -\sin(\theta_i) & 0 \\
		\sin(\theta_i) & \cos(\theta_i) & 0 \\
		0 & 0 & 0 \\
	\end{bmatrix} \frac{-\bm w_i}{\|\bm w_i\|}.
\end{eqnarray}
It follows from \eqref{eq:normvector} that the orientation of  IRS $i$ can be changed by adjusting $\theta_i$. For each candidate IRS site, we impose the following constraints on the IRS orientation:
\begin{eqnarray} \label{Cons:oritentation}
	\theta_i \in {\cal O}_i, \quad i \in {\cal I}_0,
\end{eqnarray}
where ${\mathcal O}_i \triangleq \{\tilde{\theta}_{i,1},\ldots, \tilde{\theta}_{i,Q_i} \}$  denotes the set of $Q_i$ possible discrete orientation angles at candidate IRS site $i$, $i \in {\cal I}_0$. 
For convenience, in the sequel of this paper, we let ${\bm s}_i \triangleq (h_{i},\theta_i) \in {\cal H}_i \times {\cal O}_i$ and denote by ${\cal S} \triangleq \{{\bm s}_i \, | \, i \in {\cal I}\}$ the set of  IRS deployment heights and orientations at these selected candidate sites.

In addition, similar to the approach in \cite{fu2024multi1}, to practically mount the IRS at each candidate site, we assume that each IRS is composed of a number of tiles (or sub-surfaces) of the same fixed size. Let $M$ denote the number of reflecting elements along both horizontal and vertical dimensions per tile, and $T_{i}$ represent the number of tiles on IRS $i$. Consequently, the total number of reflecting elements on IRS $i$ is given by $M_i \triangleq T_iM^2$.
Given the practical constraints on the size of IRS, we assume a maximum allowable number of tiles that can be deployed at each candidate site, denoted as $T^{\max}_0$. 
Therefore, we have $T_i \leq T^{\max}_0, i \in {\cal I}_0$. 
For each site $i$ without IRS deployed, we set $T_i = 0$.
For convenience, in the sequel of this paper, we refer to tile numbers only pertaining the candidate sites with IRSs  deployed, i.e., $i \in {\cal I}$. Accordingly, let ${\cal T} \triangleq \{T_i \, | \, i \in {\cal I}\}$ denote the set of tile numbers deployed at these candidate sites.

\begin{figure}[t]
	\centering
	\subfigure{\includegraphics[scale=0.25]{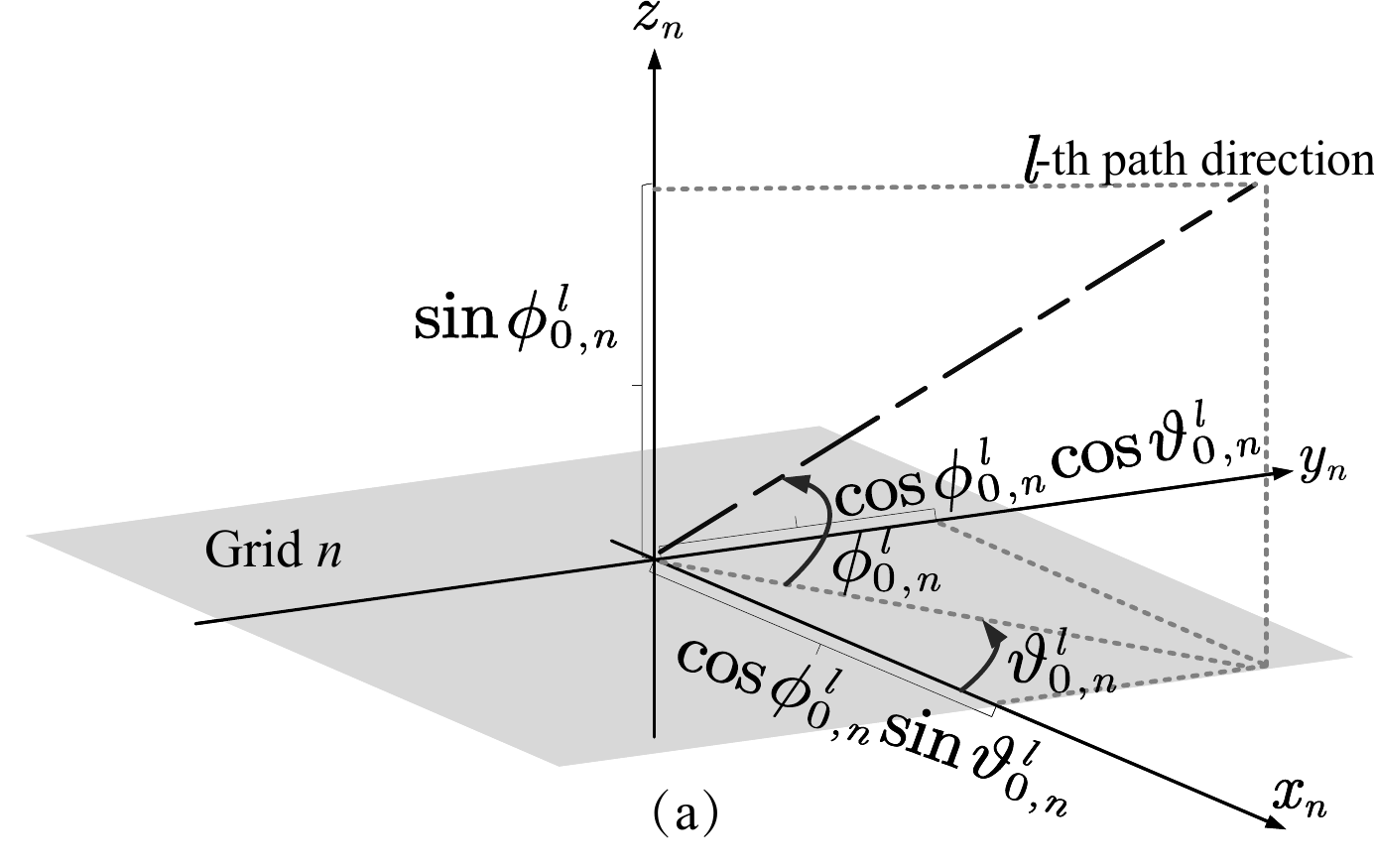}\label{Fig:SystemModel3}}
	\subfigure{\includegraphics[scale=0.25]{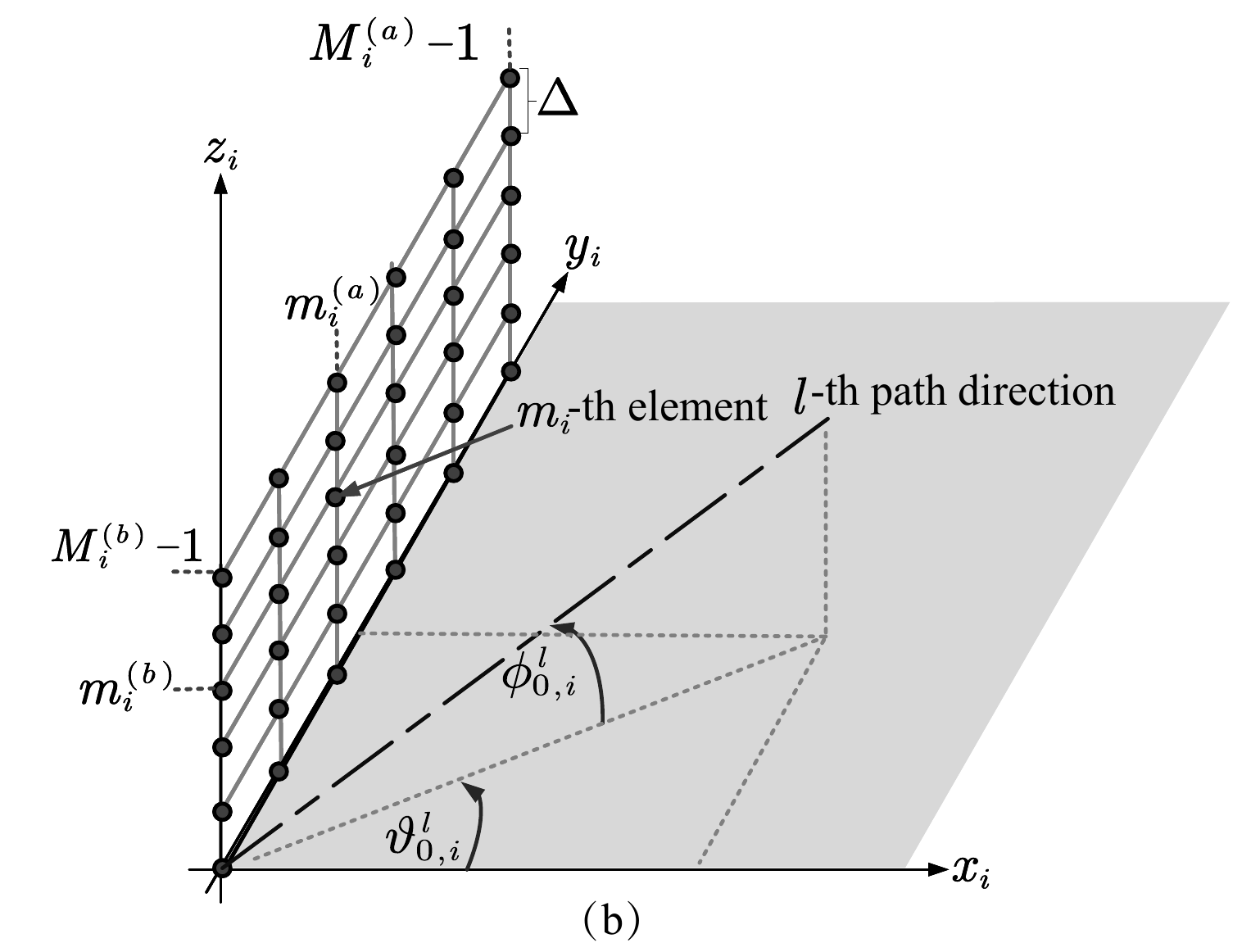}\label{Fig:SystemModel4}}	
		\vspace{-2mm}
	\caption{Illustration of the setup where a plane wave impinges on (a) grid $n$, and (b) IRS $i$. }\label{Fig:SystemModel}
	\vspace{-6mm}
\end{figure}
\subsection{Channel Model}

In this subsection, we characterize the baseband equivalent channels from the BS to each IRS $i$, from each IRS $i$ to any receiver point (position) in ${\cal D}$, and from the BS to any receiver point in ${\cal D}$.
For the channels from BS/IRS to any receiver point within $\mathcal{D}$, they are determined by the signal propagation and the position of the receiver. 
To facilitate the evaluation of communication quality within the region ${\cal D}$, we uniformly divide the area into $N$ grids, each of which is assumed to be a square with side length $\delta$. 
This grid-based division allows for a more tractable and systematic analysis of large-scale channel characteristics and the coverage performance based on them.
More specifically, by examining the average performance within each individual grid, we can identify areas with coverage gaps or signal degradation, thereby enabling the problem formulation for IRS deployment optimization. Additionally, the grid-based approach simplifies the computational complexity of solving the deployment design problem, making it easier to apply the proposed solution to large areas.

Besides, we consider the far-field wireless channel model because the size of transmit/receive region (i.e., IRS and receiver grid) in our case is generally much smaller than the signal propagation distance. 
Thus, for each channel path component, all receiver positions in the same receiver grid and all element positions on the same IRS experience the same angle of departure/angle of arrival (AoD/AoA), and an invariant amplitude of the complex path coefficient, while only the phase of the complex path coefficient varies with positions\cite{zhu2024modeling}.
For convenience, we refer to the BS and receiver grid $n$ as nodes 0 and $I_0 + n$, respectively, and define $\mathcal{N} = \{I_0 + 1, \ldots, I_0 + N\}$.

First, we characterize the direct channel between the BS and grid $n, n\in {\cal N}$, which is denoted as $h _{0,n}$.
For a uniform square region at grid $n$,  we establish a local two-dimensional (2D) coordinate system, with the $x_n$-$O_n$-$y_n$ plane representing the plane in which grid $n$ is located, and the origin $O_n$ set at the center of grid $n$, as shown in Fig. \ref{Fig:SystemModel3}.
Denote the number of  dominant channel paths from the BS to the reference point of grid $ n $, $ n \in \mathcal{N} $ (i.e., the origin $O_n$) by $ L_{0,n}$.
The path response vector (PRV) from the BS to the reference point of receiver grid $n$ (i.e., grid center) is given by
\begin{eqnarray} 
\bm{u}_{0,n} \triangleq [u_{0,n}^1, \ldots, u_{0,n}^{L_{0,n}}]^{\sf T} \in \mathbb{C}^{L_{0,n}},
\end{eqnarray}
where $u_{0,n}^l$ denotes the complex path gain of the $l$-th path from the BS to receiver grid $n$.
Furthermore, the location of any point in receiver grid \( n \) is denoted as \(\bm{r}_n = [x_n, y_n, 0]\) in its local Cartesian coordinate system.
Then, using basic geometry, the difference in signal propagation distance between position \(\bm{r}_n\) and the reference point for the \( l \)-th path is given by
\begin{eqnarray}
\rho_{0,n}^l(\bm{r}_n) = x_n \cos\phi_{0,n}^l\sin \vartheta_{0,n}^l + y_n \cos\phi_{0,n}^l \cos \vartheta_{0,n}^l,
\end{eqnarray}
where \( \phi_{0,n}^l \) and \( \vartheta_{0,n}^l \) denotes the elevation and azimuth AoAs of the \( l \)-th path, respectively.
Thus, the channel response of the \( l \)-th path at position \(\bm{r}_n\) has a phase difference of \( {2\pi \rho_{0,n}^l(\bm{r}_n)}/{\lambda} \) w.r.t. the reference point, where \( \lambda \) is the carrier wavelength.
To account for these phase differences across all \( L_{0,n} \) receive paths, we define the receive field-response vector (FRV) in the receive region of grid \( n \) from the BS as\cite{zhu2024modeling}
\begin{eqnarray}\label{Eq:directFRV}
\bar{\bm{h}}_{0,n}(\bm{r}_n) \triangleq \left[ e^{j\frac{2\pi}{\lambda} \rho_{0,n}^1(\bm{r})}, \ldots, e^{j\frac{2\pi}{\lambda} \rho_{0,n}^{L_{0,n}}(\bm{r})} \right]^{\sf T} \in \mathbb{C}^{L_{0,n}}.
\end{eqnarray}
As a result, the direct channel from the BS to any point \(\bm{r}_n\) within the receiver grid \( n \) is modeled as
\begin{eqnarray}\label{Eq:directlink}
h_{0,n}(\bm{r}_n) = \bar{\bm{h}}^{\sf H}_{0,n}(\bm{r}_n) \bm{u}_{0,n}.
\end{eqnarray}

Next, we characterize the channels between the BS and any element on IRS $i$, and between the IRS and grid $n$, which are denoted by $f_{0,i}$ and $g_{i,n}$, respectively.
Note that unlike the direct channel between BS and grid $n$, the IRS $i$-related channels depend on its deployed height $h_i$ and orientation $\theta_i$.
For a uniform planar array (UPA)-based IRS $i$ with any given height and orientation (i.e., $\bm{s}_i$),  we also establish a local 2D coordinate system parallel to the $y_i$-$z_i$ plane, where the bottom-left corner of the array is defined as the origin and reference point, as shown in Fig. \ref{Fig:SystemModel4}.
For any given $\bm s_i$, let $L_{0,i}(\bm{s}_i)$ and $L_{i,n}(\bm{s}_i)$ denote the number of paths from the BS to IRS $i$ and from IRS $i$ to receiver grid $n$, respectively.
Accordingly, we respectively denote the PRVs from the BS to the reference point of IRS $i$ and from IRS $i$ to the reference point of receiver grid $n$ as 
\begin{eqnarray}
	\bm{\sigma}_{0,i}(\bm{s}_i) &\triangleq & [\sigma_{0,i}^1, \ldots, \sigma_{0,i}^{L_{0,i}(\bm{s}_i)}]^{\sf T} \in \mathbb{C}^{L_{0,i}(\bm{s}_i)}, \\
	\bm{\omega}_{i,n}(\bm{s}_i) &\triangleq & [\omega_{i,n}^1, \ldots, \omega_{i,n}^{L_{i,n}(\bm{s}_i)}]^{\sf T} \in \mathbb{C}^{L_{i,n}(\bm{s}_i)}, 
\end{eqnarray}
where  $\sigma_{0,i}^l$ and $\omega_{i,n}^l$ denote the complex path gain of $l$-path from the BS to the reference point of IRS $i$ and from IRS $i$ to the reference point of receiver grid $n$, respectively.
We denote the index pair of any particular element $m_i$ at IRS $i$ as $(m_i^{(a)}, m_i^{(b)})$, which corresponds to $y_i$ and $z_i$ directions. 
Here, we have $m_i^{(a)} \in \{0, \ldots, M_i^{(a)}-1\}$ and $m_i^{(b)} \in \{0, \ldots, M_i^{(b)}-1\}$ with $M_i = M_i^{(a)} M_i^{(b)}$. Thus, we have a unique mapping between $m_i$ and $(m_i^{(a)}, m_i^{(b)})$, i.e.,
\begin{eqnarray}
	m_i = 1 + m_i^{(a)} + m_i^{(b)} M_i^{(a)}.
\end{eqnarray}
As a result, the coordinates of the $m_i$-th reflecting element in the local coordinate system are denoted as $\bm{t}_{m_i} = (0,m_i^{(a)} \Delta, m_i^{(b)} \Delta)$ with $\Delta$ denoting the inter-element spacing. 
Then, with any given $\bm{s}_i$,  the receive FRV of the $m_i$-th element of IRS $i$ w.r.t. the BS is given by
 \begin{eqnarray}
\!\!\!\!\! &&	\bar{\bm{f}}_{0,i}(\bm{s}_i, \bm{t}_{m_i}) \triangleq \nonumber \\
\!\!\!\!\!	&& \left[ e^{j\frac{2\pi}{\lambda} \rho_{0,i}^1(\bm{s}_i, \bm{t}_{m_i})}, \ldots, e^{j\frac{2\pi}{\lambda} \rho_{0,i}^{L_{0,i}}(\bm{s}_i, \bm{t}_{m_i})} \right]^{\sf T} \in \mathbb{C}^{L_{0,i}(\bm{s}_i)},
\end{eqnarray}
where $\rho_{0,i}^l(\bm{s}_i, \bm{t}_{m_i}) = m_i^{(a)} \Delta \cos\phi_{0,i}^l \cos\vartheta_{0,i}^l + m_i^{(b)} \Delta  \sin\phi_{0,i}^l$ with \( \phi_{0,i}^l \) and \( \vartheta_{0,i}^l \) denoting the elevation and azimuth AoAs of the \( l \)-th path from the BS to IRS $i$, respectively.
Similarly, we can define transmit and receive FRVs of the $m_i$-th element of IRS $i$ and receive grid $n$ as $\bar{\bm{h}}_{i,n}(\bm{s}_i, \bm{r}_n)$ and $\bar{\bm{g}}_{i,n}(\bm{s}_i, \bm{t}_{m_i})$, respectively, as similarly to \eqref{Eq:directFRV}.

 Based on above, the channel from the BS to the $m_i$-th element of IRS $i$ is given by
\begin{eqnarray}\label{Eq:BS-IRSlink}
	f_{0,i}(\bm{s}_i, \bm{t}_{m_i}) = \bar{\bm{f}}^{\sf H}_{0,i}(\bm{s}_i, \bm{t}_{m_i}) \bm{\sigma}_{0,i}(\bm{s}_i).
\end{eqnarray}
Similarly, the channel from the $m_i$-th element on IRS $i$ to any point $\bm{r}_n$ within grid $n$ is given by
\begin{equation}\label{Eq:IRSuserlink}
	g_{i,n}(\bm{s}_i, \bm{t}_{m_i}, \bm{r}_n) = \bar{\bm{h}}^{\sf H}_{i,n}(\bm{s}_i, \bm{r}_n) \text{diag}(\bm{\omega}_{i,n}(\bm{s}_i)) \bar{\bm{g}}_{i,n}(\bm{s}_i, \bm{t}_{m_i}).
\end{equation}

In practice, the PRVs  and FRVs of all BS-grid, BS-IRS, IRS-grid links can be acquired by using, e.g., ray tracing \cite{fuschini2015ray} or channel measurement.
However, even with such information available, it remains challenging to characterize the coverage performance of each receiver grid based on the exact channel state information (CSI) due to the infinite number of possible receiver locations within each receiver grid $n$. Moreover, the effect of multi-path small-scale fading complicates the design of IRS deployment to simultaneously enhance the signal power for all possible receiver locations within the same grid.
To address these issues, the primary objective of this paper is to strategically deploy IRSs by leveraging the knowledge on large-scale CSI, where the average channel power gain over all locations within each grid is adopted as the performance measurement so as to eliminate the effect of multi-path small-scale fading in it. Building upon the channel models described above, we will first derive the average channel power gains from the BS to each IRS/receiver grid and from each IRS to each receiver grid, respectively, and then approximate the average channel power gains of the cascaded BS-IRS-grid links in the following subsection.

\subsection{Large-Scale Channel Power Characterization}\label{Subsection:large-scale}
First, the average BS-grid $n$ channel power gain can be obtained by calculating the average value of the squared magnitude of  $h_{0,n}(\bm{r}_n)$ in \eqref{Eq:directlink} over grid $n$, which is given by
\begin{eqnarray}\label{Eq:integral11}
  &&\mathbb{E}_{\bm{r}_n}\left[ |h_{0,n}(\bm{r}_n)|^2 \right]   = \frac{1}{\delta^2}\int_{-\delta/2}^{\delta/2}\int_{-\delta/2}^{\delta/2}|h_{0,n}(\bm{r}_n)|^2dx_ndy_n \nonumber\\
	&  = & \sum_{l = 1}^{L_{0,n}}|u^l_{0,n}|^2 +
	 2\sum_{l = 1}^{L_{0,n}}\sum_{l' < l}|u^l_{0,n}u^{l'}_{0,n}|\varrho_{l,l'}, 
\end{eqnarray}
with 
\begingroup\makeatletter\def\f@size{9.2}\check@mathfonts\def\maketag@@@#1{\hbox{\m@th\normalsize\normalfont#1}}
\begin{eqnarray} \label{Eq:integral12}
 &&\!\!\!\! \rho_{l,l'} = \frac{1}{\delta^2}\int_{-\delta/2}^{\delta/2}\int_{-\delta/2}^{\delta/2}\cos(\frac{2\pi}{\lambda}(\rho_{0,n}^l(\bm{r}_n)-\rho_{0,n}^{l'}(\bm{r}_n)))dx_ndy_n\nonumber \\
 &&\!\!\!\! = \!\!\frac{\lambda^2 \!\left[ \cos\left(\frac{2\pi (A_{l,l'} \!+\! B_{l,l'})\delta}{\lambda}\right) \!-\! \cos\left(\frac{2\pi A_{l,l'}\delta}{\lambda}\right) \!-\! \cos\left(\frac{2\pi B_{l,l'}\delta}{\lambda}\right) \!+\! 1 \!\right]}{4 \pi^2 \delta^2 A_{l,l'} B_{l,l'}}\nonumber \\
 && \leq \frac{\lambda^2}{\pi^2 \delta^2 A_{l,l'} B_{l,l'}},
\end{eqnarray}\normalsize
where  $A_{l,l'} = \cos\phi_{0,n}^l\sin \vartheta_{0,n}^l-\cos\phi_{0,n}^{l'}\sin \vartheta_{0,n}^{l'}$ and $B_{l,l'} = \cos\phi_{0,n}^{l}\cos \vartheta_{0,n}^{l} - \cos\phi_{0,n}^{l'}\cos \vartheta_{0,n}^{l'}$.
We can observe from \eqref{Eq:integral12} that $\rho_{l,l'}$ is upper-bounded by $\frac{\lambda^2}{\pi^2 \delta^2 A_{l,l'} B_{l,l'}}$, since each cosine term is upper-bounded by one.
Furthermore,  if the angles $\phi_{0,n}^l$, $\phi_{0,n}^{l'}$, $\vartheta_{0,n}^l$, and $\vartheta_{0,n}^{l'}$ are different from each other such that their sine and cosine terms lead to $A_{l,l'}$ and $B_{l,l'}$ being non-zero,
each term $\rho_{l,l'}$ in \eqref{Eq:integral11} will approach zero because the size of each grid $\delta$ is considered to be much larger than the wavelength, i.e., $\delta \gg \lambda$.
Therefore,  $\mathbb{E}_{\bm{r}_n}\left[ |h_{0,n}(\bm{r}_n)|^2 \right]$ can be approximated as 
\begin{eqnarray} \label{Eq:approx1}
\mathbb{E}_{\bm{r}_n}\left[ |h_{0,n}(\bm{r}_n)|^2 \right]  \approx  \|\bm u_{0,n}\|^2  \triangleq  \beta_{0,n}.
\end{eqnarray}
Such an approximation characterizes the impact of dominant channel paths from the BS to any receiver grid on the large-scale channel power gain averaged over any grid $n$, which will facilitate our subsequent IRS deployment design.

Similarly, for any given $h_i$ and $\theta_{i}$ (i.e., $\bm{s}_i$), the average channel power gain between the BS and any element on IRS $i$ is the value of the squared magnitude of  $f_{0,i}(\bm{s}_i, \bm{t}_{m_i})$ in \eqref{Eq:BS-IRSlink} averaged over all its  reflecting elements, which is given by
\begin{eqnarray}\label{Eq:approx21}
	 &&\mathbb{E}_{\bm{t}_{m_i}}\left[ |f_{0,i}(\bm{s}_i, \bm{t}_{m_i})|^2 \right] = \frac{\sum_{m_i=1}^{M_i}|f_{0,i}(\bm{s}_i, \bm{t}_{m_i})|^2}{M_i} \nonumber \\
	 &=& \sum_{l = 1}^{L_{0,i}}|\sigma^l_{0,i}|^2 \!+\! 2 \sum_{l = 1}^{L_{0,n}}\sum_{l' < l}|\sigma^l_{0,i}\sigma^{l'}_{0,i}|\tilde{\rho}_{l,l'}, 
\end{eqnarray}
with
\begingroup\makeatletter\def\f@size{9.2}\check@mathfonts\def\maketag@@@#1{\hbox{\m@th\normalsize\normalfont#1}}
\begin{eqnarray} \label{Eq:approx22}
	\tilde{\rho}_{l,l'} &=& \frac{\sum_{m_i^{a}=1}^{M_i^{a}}\sum_{m_i^{b}=1}^{M_i^{b}}\cos(\frac{2\pi}{\lambda}(\rho_{0,i}^l(\bm{s}_i, \bm{t}_{m_i})-\rho_{0,i}^{l'}(\bm{s}_i, \bm{t}_{m_i})))}{M_i^{a}M_i^{b}} \nonumber\\
&\leq& 
	\frac{1}{M_i^{a} M_i^{b}} \frac{\sin\left(M_i^{a}\frac{\pi}{\lambda} \Delta  \tilde{A}_{l,l'}  \right) \sin\left(M_i^{b}\frac{\pi}{\lambda} \Delta  \tilde{B}_{l,l'}   \right)}{\sin\left(\frac{\pi}{\lambda} \Delta  \tilde{A}_{l,l'} \right) \sin\left( \frac{\pi}{\lambda} \Delta  \tilde{B}_{l,l'} \right)},
\end{eqnarray}\normalsize
where 
$\tilde{A}_{l,l'} = \cos\phi_{0,i}^l \sin\vartheta_{0,i}^l - \cos\phi_{0,i}^{l'} \sin\vartheta_{0,i}^{l'}$ and $\tilde{B}_{l,l'} = \sin\phi_{0,i}^l - \sin\phi_{0,i}^{l'}$.
Similar to the derivation in \eqref{Eq:integral12}, if all channel paths are sufficiently distinguishable in the angular domain, each term $\tilde{\rho}_{l,l'}$ in \eqref{Eq:approx21} will approach zero, as $M_i^{a} M_i^{b}$ (i.e., the size of IRS $i$) is considered to be much larger than one. 
Therefore, in the following, we approximate $\mathbb{E}_{\bm{t}_{m_i}}\left[ |f_{0,i}(\bm{s}_i, \bm{t}_{m_i})|^2 \right]$  as 
\begin{eqnarray} \label{Eq:approx2}
	\mathbb{E}_{\bm{t}_{m_i}}\left[ |f_{0,i}(\bm{s}_i, \bm{t}_{m_i})|^2 \right]  \approx\|\bm{\sigma}_{0,i}(\bm{s}_i)\|^2.
\end{eqnarray}
Similarly, for any given $h_i$ and $\theta_{i}$, the average channel power gain between any element on the IRS to grid $n$ is calculated by the average value of the squared magnitude of the function $g_{i,n}(\bm{s}_i, \bm{t}_{m_i}, \bm{r}_n)$ in \eqref{Eq:IRSuserlink} over all its reflecting elements and/or all points over grid $n$ and thus approximated as
\begin{eqnarray} \label{Eq:approx3}
	\mathbb{E}_{\bm{r}_{n}}\left[\mathbb{E}_{\bm{t}_{m_i}}\left[ |g_{i,n}(\bm{s}_i, \bm{t}_{m_i}, \bm{r}_n)|^2 \right] \right] \approx\|\bm{\omega}_{0,i}(\bm{s}_i)\|^2.
\end{eqnarray}

Finally, for any given $\bm s_i$ and $T_i$, we calculate the  average  channel power gain of  the cascaded  BS-IRS $i$-grid $n$ link, which is denoted by $|\tilde{h}_{0,n}(\bm s_i,T_i)|^2$.
 In practice, the phase-shift matrix of each IRS should be designed the actual channels of users at specific locations in different girds to maximize the passive reflection gain, which is upper-bounded by $T_i^2M^4$ (including the aperture gain $T_iM^2$ and beamforming gain $T_iM^2$) for the case of LoS link \cite{Wu2021Tutorial}. However, for our considered multi-path channels, achieving the upper bound on beamforming gain is generally not feasible, and designing the phase-shift matrix of each IRS to obtain the maximum reflection gain for each user at any arbitrary location is analytically intractable.
 To make the IRS deployment problem  tractable and our proposed solution robust to users' spatially varying channels within each grid,
 we resort to the approximated average BS-IRS  $i$-grid $n$ cascaded channel power gain, without needing the sophisticated phase shift design for IRSs based on user's exact channels at different locations.
 To this end, we define an effective reflection gain of each IRS to characterize its average power amplification contributed to the cascaded channel. 
 On one hand, IRS $i$ can achieve an inherent aperture gain of order $T_iM^2$ by collecting signal power in the link from the BS to IRS $i$. 
 On the other hand, its reflection gain needs to be allocated to $L_{0,i}(\bm{s}_i)L_{i,n}(\bm{s}_i)$ pairs of incident and reflected channel paths for improving the average channel power, which yields an effective power gain of IRS $i$ as $\frac{T_iM^2}{L_{0,i}(\bm{s}_i)L_{i,n}(\bm{s}_i)}$. Thus, the overall effective reflection gain of IRS $i$ is given by  $T_iM^2 \times \frac{T_iM^2}{L_{0,i}(\bm{s}_i)L_{i,n}(\bm{s}_i)} = \frac{T_i^2M^4}{L_{0,i}(\bm{s}_i)L_{i,n}(\bm{s}_i)}$ and  the average channel power gain of the cascaded BS-IRS $i$-grid $n$ link can be approximated by
 \begin{eqnarray}\label{Eq:IRS-reflectionGain}
 	|\tilde{h}_{0,n}(\bm s_i, T_i)|^2
 	= T_i^2M^4 \kappa_{0,i,n}(\bm s_i),
 \end{eqnarray}
where $\kappa_{0,i,n}(\bm s_i) \triangleq \|\bm{\sigma}_{0,i}(\bm{s}_i)\|^2\|\bm{\omega}_{i,n}(\bm{s}_i)\|^2/(L_{0,i}(\bm s_i)L_{i,n}(\bm s_i))$.
Note that this approximation avoids the complicated phase shift design for each IRS while efficiently capturing its essential reflection gain in practice, which greatly simplifies the multi-IRS deployment optimization as shown in the next section.

\section{Multi-IRS Deployment Problem Formulation}\label{Sec_formulation}

In this section, we first characterize the total cost for IRS deployment and the coverage performance in our considered system, and then introduce the problem formulation for multi-IRS deployment optimization.

\subsection{Deployment Cost}

We consider that the total IRS deployment cost depends on both the number of sites selected for IRS deployment and the number of tiles deployed at these selected sites \cite{fu2024multi1}. Accordingly, we express the total deployment cost as 
\begin{align}\label{Eq:deploy cost}
	c({\cal I}, {\cal T}) =  \underbrace{c_{s, 0} |{\cal I}|}_{\text{Site-use cost}}+ \underbrace{  c_{h,0}\sum\nolimits_{i \in {\cal I}}T_{i}  }_{\text{Hardware cost}},
\end{align}
where $c_{s,0}$  denotes a fixed cost for deploying an IRS at any candidate site, and $c_{h,0}$ denotes the cost per IRS tile.
Importantly note that $c_{s,0}$ captures the site-use cost of deploying IRS, including mounting cost, control cost, and other related expenses, which are independent of the size of the IRS (or the number of tiles deployed). In contrast, $c_{h,0}$ captures the hardware cost of deploying an IRS tile, which can be determined based on its manufacturing cost and power consumption.

\subsection{Coverage Performance Metric}

By strategically deploying IRSs within ${\mathcal D}$, both direct and IRS-reflection links can be established from the BS to each receiver grid and the desired user locations within it, thereby enhancing the received signal power. Consequently, this paper considers the sum of the average channel power gains of all possible direct and single-IRS-reflection links from the BS to each receiver grid as a measure of channel quality, referred to as the effective channel power gain. 
This average channel power performance is influenced by the number and locations of sites deployed with IRSs (i.e., ${\mathcal I}$) as well as their associated heights, orientations, and tile numbers (i.e., $\mathcal S$ and ${\mathcal T}$). 
Deploying IRSs at more sites can create additional signal paths, thereby enhancing the coverage of the BS. 
For any given ${\mathcal I}$ and ${\mathcal T}$, as shown in \eqref{Eq:IRS-reflectionGain}, optimizing each IRS's orientation and height can effectively change the number of channel paths over its front half-space, thus improving IRS-related path gains. 
Furthermore, for any given ${\mathcal I}$ and ${\mathcal S}$, increasing the number of tiles per IRS can enhance the signal strength over each single-IRS-reflection path due to the improved effective reflection gain, as observed in \eqref{Eq:IRS-reflectionGain}.

It follows from \eqref{Eq:approx1} and \eqref{Eq:IRS-reflectionGain} that, for any given ${\mathcal I}$, ${\mathcal S}$, and ${\mathcal T}$, the overall average channel power gain (i.e., the effective channel power gain) of both the direct BS-grid $n$ link and the cascaded BS-IRS-grid $n$ link via all the IRSs in ${\cal I}$ is given by 
\begin{eqnarray}\label{Signalpower}
	\bar{P}_{n}(\mathcal{I},{\mathcal S}, {\mathcal T})  
	&=  &  \beta_{0,n} +  \sum_{i\in\mathcal{I}} 	|\tilde{h}_{0,n}(\bm s_i, T_i)|^2\nonumber \\
	&= &
	\beta_{0,n} + \sum_{i\in\mathcal{I}} T_i^2M^4\kappa_{0,i,n}(\bm s_i), n\in{\mathcal N}.
\end{eqnarray}
Furthermore, an outage event occurs for grid $n$ when its effective channel power gain in \eqref{Signalpower} is lower than a minimum threshold $P^{\min}$. 
Accordingly, for any given ${\mathcal I}$, ${\mathcal S}$, and ${\mathcal T}$, the coverage rate among $N$ receiver grids is defined as 
\begin{eqnarray}\label{Eq:objective}
	\eta(\mathcal{I},{\mathcal S}, {\mathcal T}) &=& \frac{\sum_{n\in{\mathcal N}}u(\bar{P}_{n}(\mathcal{I},{\mathcal S}, {\mathcal T}))-P^{\min} )}{N}, 
\end{eqnarray}
where $u(x)$ is the step function (i.e., $u(x)=1$ if $x\geq 0$; otherwise, $u(x)=0$). On the right-hand side of \eqref{Eq:objective}, the numerator represents the number of receiver grids whose effective channel power gains  meet the requirement $P^{\min}$.
It should be noted that the coverage rate defined in \eqref{Eq:objective} is used to characterize the long-term coverage performance over the region of interest (i.e., ${\mathcal D}$) based on the average channel power gains, without any users' instantaneous CSI available at the stage of IRS deployment. In this paper, we focus  on the IRS deployment problem based on the large-scale channel knowledge and do not consider the design of IRS reflection coefficients based on any specific instantaneous CSI. Nevertheless, the channel estimation and IRS reflection design can be executed after  IRSs have been deployed at selected sites in ${\cal D}$, which is beyond the scope of this paper\footnote{Interested readers can refer to  existing works (e.g., \cite{FU2021Reconfigurable, yan2024power,sun2024power} on the other IRS design aspects.}.

\subsection{Problem Formulation}\label{Sec_Formulation2}

In this paper, we aim to jointly optimize the IRS deployment locations/sites (i.e., ${\cal I}$), as well as their associated heights, orientations, and tile numbers (i.e., ${\cal S}$ and ${\cal T}$), to minimize the total deployment cost as defined in \eqref{Eq:deploy cost}, subject to constraints on the coverage rate performance in \eqref{Eq:objective} and practical IRS deployment considerations. 
The associated optimization problem is formulated as
\begin{subequations}
	\begin{eqnarray} \label{Problem1}
		\text{(P0):} \mathop{\text{Minimize}}\limits_{{\cal I}, {\cal T},{\mathcal S}} && c({\cal I}, {\cal T}) \nonumber\\
		\text{subject to}\;\;&&{\cal I} \subseteq {\cal I}_0, \label{Cons: Location2}\\
		&& 	\eta(\mathcal{I},{\mathcal S}, {\mathcal T}) \geq \eta_0,  \label{Cons:coveragerate}\\  
		&& {\bm s}_i \in {\cal H}_i \times {\cal O}_i, \forall i \in {\cal I},  \label{Cons:State} \\
		&& 	T_i \leq T_0^{\max}, T_i \in \mathbb{N}^{+}, \forall i \in {\cal I}, \label{Cons:Tile}
\end{eqnarray}	
\end{subequations}
where $\eta_0 \leq 1$ denotes a prescribed coverage rate target. 
We first analyze the feasibility of (P0). For the special case that no IRS is allowed to be deployed, i.e., ${\cal I} = \emptyset$, $\cal S = \emptyset$, and ${\cal T} = \emptyset$, the achievable coverage rate in \eqref{Eq:objective} is given by  $\eta^{\min } \triangleq \frac{\sum_{n\in{\mathcal N}}u(\beta_{0,n}^2-P^{\min} )}{N}$.
If $\eta_0\leq \eta^{\min }$, (P0) is always feasible with a zero-value objective.
To render (P0) non-trivial, this paper considers cases where $\eta_0> \eta^{\min }$ for a given $P^{\min }$.  
On the other hand, for any given $P^{\min }$, a maximum $\eta_0$ should exist to ensure that (P0) is feasible, due to the maximum tile number constraints in \eqref{Cons:Tile}. 
In particular, for any grid $n$, the maximum value of effective channel power gain $\bar{P}_{n}(\mathcal{I},{\mathcal S}, {\mathcal T})$ in \eqref{Signalpower} can be achieved by deploying IRSs at all candidate sites in ${\cal I}_0$, each equipped with the maximum number of tiles, while optimizing their heights and orientations to maximize each term $\kappa_{0,i,n}(\bm{s}_i)$ in \eqref{Signalpower}, i.e., ${\cal I}^{\max} = {\cal I}_0$, ${\cal T}^{\max} = \{T_{0}^{\max}  |  i \in {\cal I}_0\}$, and ${\cal S}_n^{\max} = \{\arg\max_{{\bm s}_i \in {\cal H}_i \times {\cal O}_i}\kappa_{0,i,n}({\bm s}_i)  |  i \in {\cal I}_0\}$. 
However, the optimal height and orientation solutions ${\cal S}_n^{\max}$ for different grids $n$ may be distinct.
Therefore, for any given $P^{\min }$, an achievable $\eta_0$ is upper-bounded by $\frac{\sum_{n\in{\mathcal N}}u(\bar{P}_{n}({\cal I}^{\max},{\cal S}_n^{\max}, {\cal T}^{\max}))-P^{\min})}{N}$.

It is important to note that (P0) is a combinatorial optimization problem, which is generally challenging to solve optimally using conventional convex optimization algorithms. To address this issue, in Section \ref{Sec:Solution}, we will first reformulate (P0) as an ILP problem, which can be optimally solved using the BB algorithm \cite{narendra1977branch}. 
To further reduce computational complexity, we will also propose an efficient successive refinement method to solve (P0).

\vspace{-2mm}
\section{Proposed Solutions}\label{Sec:Solution}

In this section, we first present the optimal solution to (P0) via the BB algorithm, followed by a lower-complexity successive refinement method to solve (P0) more efficiently.

\vspace{-2mm}
\subsection{Optimal  Deployment by the BB Algorithm} \label{OptimalOne}
Before applying the BB algorithm to solve (P0), the problem must be reformulated into a linear programming program, which involves the following key operations.
 
First, we introduce a set of binary variables, $\bm \Xi \triangleq \{\xi^{i}_{tjk}\}$, where $i \in {\cal I}_0$, $t\in \{1,\ldots,  T_{0}^{\max}\}$, $ j \in \{1,\ldots,  |{\cal H}_i|\}$, $k \in \{1,\ldots,  |{\cal O}_i|\}$, to represent potential deployment configurations (i.e., height, orientation, and tile number) at any candidate IRS site $i$. 
Specifically,  $\xi^{i}_{tjk} = 1$ if candidate site $i$ is selected for deploying an IRS with tile number $t$, height $\tilde{h}_{i,j}$, and orientation $\tilde{\theta}_{i,k}$; otherwise,  $\xi^{i}_{tjk} = 0$.
Given that each candidate IRS site is allowed to deploy at most one IRS, the following linear constraints must be imposed on variables $\{\xi^{i}_{tjk}\}$:
\begin{eqnarray} \label{Cons:binaryvariable1}
	\sum_{t = 1}^{T_{0}^{\max}}\sum_{j = 1}^{|{\cal H}_i|}\sum_{k = 1}^{|{\cal O}_i|} \xi^{i}_{tjk} \leq 1, \forall i.
\end{eqnarray}
These constraints ensure that for any $i$, an IRS should be deployed at candidate site $i$ if the equality holds; otherwise, no IRS is deployed at that site.

Consequently, the deployment cost $c(\mathcal{I},{\mathcal T})$ in \eqref{Eq:deploy cost} can be reformulated into a linear form w.r.t. $\bm \Xi$:
\begin{eqnarray}\label{Eq:deploycost2}
	c(\bm \Xi) =  \sum_{i \in {\cal I}_0}\sum_{t = 1}^{T_{0}^{\max}}\sum_{j = 1}^{|{\cal H}_i|}\sum_{k = 1}^{|{\cal O}_i|} (c_{s, 0} + c_{h,0}t) \xi^{i}_{tjk}.		
\end{eqnarray}

Furthermore, we reformulate the coverage rate constraint in \eqref{Cons:coveragerate} to achieve a linear representation, where the most challenging issue is how to deal with  $\eta(\mathcal{I},{\mathcal S}, {\mathcal T})$.
Initially, $\bar{P}_{n}(\mathcal{I},{\mathcal S}, {\mathcal T})$ in \eqref{Signalpower}, which is involved in $\eta(\mathcal{I},{\mathcal S}, {\mathcal T})$, can be expressed in a linear form w.r.t. $\bm \Xi$.
In particular, $|\tilde{h}_{0,n}(\bm s_i, T_i)|^2$ in \eqref{Eq:IRS-reflectionGain} can be rewritten as
\begin{eqnarray}
		|\tilde{h}_{0,n}(\bm \Xi)|^2 =
	\sum_{t = 1}^{T_{0}^{\max}}\sum_{j = 1}^{|{\cal H}_i|}\sum_{k = 1}^{|{\cal O}_i|} \alpha_{tjk}^{in} \xi^{i}_{tjk},
\end{eqnarray}
where $\alpha_{tjk}^{in}\triangleq t^2M^2\kappa_{0,i,n}(\tilde{h}_{i,j}, \tilde{\theta}_{i,k})$.
Thus, $\bar{P}_{n}(\mathcal{I},{\mathcal S},{\mathcal T})$ in \eqref{Signalpower} can be rewritten as
\begin{eqnarray}\label{Signalpower2}
	\bar{P}_{n}(\bm \Xi) 	 =   \beta_{0,n} 
	+ \sum_{i\in\mathcal{I}_0}\sum_{t = 1}^{T_{0}^{\max}}\sum_{j = 1}^{|{\cal H}_i|}\sum_{k = 1}^{|{\cal O}_i|} \alpha_{tjk}^{in} \xi^{i}_{tjk}. 
\end{eqnarray}

However,  given the non-linear step function involved in $\eta(\mathcal{I},{\mathcal S}, {\mathcal T})$, it cannot be directly expressed as a linear function w.r.t. $\bm \Xi$ alone.
To address this issue, we introduce an additional set of binary variables,  $\bm \chi \triangleq \{\chi_n\}$, where $\forall n \in {\cal N}$, to indicate whether $\bar{P}_{n}(\bm \Xi) \geq P^{\min}$ holds. Each of these variables is subject to the following constraint:
\begin{eqnarray}\label{Cons:coverIndicator}
	P^{\min}\chi_n- \bar{P}_{n}(\bm \Xi) &\leq& 0, \forall n \in {\cal N}. 
\end{eqnarray}
Constraint \eqref{Cons:coverIndicator} ensures that for any $n$,
 $\chi_n = 0$ if $\bar{P}_{n}(\bm \Xi) < P^{\min}$; otherwise, $\chi_n$ can take a value of 1 or 0.
Consequently, $\eta(\mathcal{I},{\mathcal S}, {\mathcal T})$ can be expressed in a linear form w.r.t.  $\bm \chi$ as 
\begin{eqnarray}
	\eta(\bm \chi) &=& \frac{\sum_{n\in {\cal N}}\chi_n}{N}. 
\end{eqnarray}

Based on the above introduced variables and constraints, problem (P0) can be equivalently transformed into the following ILP problem,
\begin{subequations}
\begin{eqnarray} \label{Problem2}
	\text{(P1):} \mathop{\text{Minimize}}\limits_{\bm \Xi, \bm \chi} && \sum_{i  \in {\cal I}_0}\sum_{t = 1}^{T_{0}^{\max}}\sum_{j = 1}^{|{\cal H}_i|}\sum_{k = 1}^{|{\cal O}_i|} (c_{s, 0} + c_{h,0}t) \xi^{i}_{tjk}  \nonumber\\
	\text{subject to}\;\;
	&& \frac{\sum_{n\in {\cal N}}\chi_n}{N}\geq \eta_0, \label{Cons:CoverRate}  \\   
    && \text{Constraints}~\eqref{Cons:binaryvariable1}~,\eqref{Cons:coverIndicator},\\
    && \xi^{i}_{tjk}\in \{0,1\}, \forall i \in {\cal I}_0, \forall t, \forall j,  \forall k,\\
    && \chi_n \in \{0,1\}, \forall n \in {\cal N}.
\end{eqnarray}	
\end{subequations}
Problem (P1) involves $(N + \sum_{i=1}^{I_0}T_0^{\max} |{\cal H}_i| |{\cal O}_i|)$ binary variables and $(1+N + I_0)$ linear inequality constraints.
Consequently, this problem can be optimally solved using the BB algorithm \cite{narendra1977branch}, which systematically solves a sequence of linear programming problems.
However, it is challenging to precisely quantify the computational complexity of the BB algorithm for solving (P1), while it generally increases with the number of variables, such as $I_0$ and $T_0^{\max}$.
In the worst case, the computational complexity of the BB algorithm approaches that of full enumeration, with computational complexity in the order of $2^{(N + \sum_{i=1}^{I_0}T_0^{\max} |{\cal H}_i| |{\cal O}_i|)}$.
However, the BB algorithm typically requires significantly less running time than full enumeration, as it effectively prunes solution sets that cannot yield the optimal solution to (P1).

\begin{algorithm}[t]
	\caption{Sequential IRS Deployment Algorithm for (P2)}\label{Alg:1}
	\begin{algorithmic}[1]
		\STATE \textbf{Input:} ${\cal I}_0^{*}$.
		\STATE \textbf{Output:} ${\cal I}$, ${\cal S}$, ${\cal T}$.

		\STATE Initialize ${\cal I} = \emptyset$, ${\cal S}= \emptyset$, and ${\cal T} = \emptyset$.
		
		\WHILE{$\eta(\mathcal{I},{\mathcal S}, {\mathcal T}) < \eta_0$ and $|\mathcal{I}| < |{\cal I}_0^*|$}
		
		\STATE  For each $i\in {\cal I}_0^*\setminus \mathcal{I}$, compute $T_{i}^*$ and $\bm s_{i}^*$ using \eqref{Eq:tileNumber} and \eqref{Eq:State}, respectively.

	\STATE Define  $\eta_{i} \triangleq \eta(\mathcal{I}\cup \{ i\}, {\mathcal S} \cup \{\bm s_{i}^*\}, {\mathcal T}\cup \{ T_{i}^*\})$, for each $i\in {\cal I}_0^*\setminus \mathcal{I}$.
	\STATE Let ${\cal C} \triangleq  \{i \mid \eta_i \geq \eta_0, i\in  {\mathcal I}_0^{*} \setminus {\mathcal I}\}$.

		\IF {${\cal C}\neq \emptyset$ }
		\STATE Sort all elements in ${\cal C}$ primarily by their $T_i$'s values (in an ascending order) and secondarily by their $\eta_i$'s values (in a descending order), denoted as $\{\pi_{1},\ldots, \pi_{|\cal C|}\}$.
	
		\ELSE
		\STATE  Sort all elements in ${\cal I}_0^*\setminus \mathcal{I}$ primarily by their $\eta_i$'s values (in a descending order) and secondarily by their $T_i$'s values (in an ascending order), denoted as $\{\pi_{1},\ldots, \pi_{|{\cal I}_0^*\setminus \mathcal{I}|}\}$.
		
		\ENDIF		
		\STATE Let $i^{*} = \pi_{1}$. And update ${\cal I} \leftarrow {\cal I}\cup \{i^{*}\}$, ${\cal S} \leftarrow {\cal S}\cup \{s_{i^{*}}^{*}\}$, and ${\cal T}\leftarrow {\cal T} \cup \{T_{i^{*}}^{*} \}$.
		\ENDWHILE					
	\end{algorithmic}
\end{algorithm}

\vspace{-1em}
\subsection{Suboptimal Deployment by the Successive Refinement Algorithm}

To address the computational challenges associated with the BB algorithm, we propose a more efficient successive refinement algorithm for solving (P0) in this subsection. 
This suboptimal algorithm can significantly reduce the running time compared to the proposed optimal BB algorithm while still achieving near-optimal solutions, as will be demonstrated in Section \ref{Sec_Results}. 
The basic idea of this approach is to iteratively solve a sequence of subproblems of (P0) by replacing ${\cal I}_0$  in \eqref{Cons: Location2} with successively refined and reduced subsets of  ${\cal I}_0$, until no further reduction in deployment cost is achievable.
Additionally, to solve each subproblem of (P0), we propose a sequential IRS deployment algorithm with a polynomial complexity, which sequentially selects  the highest-priority candidate site for deploying IRS and determines its deployed tile number, height, and orientation in closed form.

\begin{algorithm}[t]
	\caption{Successive Refinement Method for Solving (P0)} \label{Alg:2}
	\begin{algorithmic}[1]
		\STATE \textbf{Input:} ${\cal I}_0$.
		\STATE \textbf{Output:} ${\cal I}$, ${\cal S}$, ${\cal T}$, and $c({\cal I},{\cal S},{\cal T})$. 
		
		\STATE Initialize ${\cal I}_0' = {\cal I}_0$,  and $c_0 = +\infty$. 
		
		\STATE	\textbf{Step 1}:   Solve (P2.1) to obtain ${\cal I}_0^{*}$ in \eqref{eq:subset} and then solve (P2) via Algorithm \ref{Alg:1} to obtain $ \{{\cal I}^{*}, {\cal S}^{*}, {\cal T}^{*}\}$.
		
		\STATE \textbf{Step 2:} 
		\STATE Update ${\cal I}_0^{*}  = {\cal I}^{*}$;
		\STATE Let $m = 1$;
		\WHILE{$m\leq |{\cal I}_0^{*}|$}
		
		\STATE Let  $m'=1$ and ${\cal I}_0' = {\cal I}_0\setminus {\cal I}_0^{*}|$
		\WHILE{$m'\leq |{\cal I}_0' |$}		
		\STATE Fix $i$ as the $m'$-the candidate site in ${\cal I}_0'$.
		\STATE Solve (P2.2) via Algorithm \ref{Alg:1} and obtain $ \{\tilde {\cal I}^{*}, \tilde{\cal S}^{*}, \tilde{\cal T}^{*}\}$.
		\IF {$ c(\tilde {\cal I}^{*}, \tilde{\cal T}^{*}) < c({\cal I}^{*}, {\cal T}^{*})$}
		\STATE   Update $\{{\cal I}^{*}, {\cal S}^{*}, {\cal T}^{*}\}\leftarrow \{\tilde {\cal I}^{*}, \tilde{\cal S}^{*}, \tilde{\cal T}^{*}\}$.
		\STATE   Update ${\cal I}_0^{*}$ by updating $i_{m}$ as $i$.
		\ENDIF	
		\STATE Update $m' = m'+1$.
		\ENDWHILE
		\STATE Update $m = m+1$.
		\ENDWHILE		
		\STATE	\textbf{Step 3}: 
		 Update ${\cal I}_0^{*}  = {\cal I}^{*}$ and solve (P2.3) via the BB algorithm to obtain solution  $\{{\cal I}, {\cal S}, {\cal T}\}$ of (P0).			
	\end{algorithmic}
\end{algorithm}

\subsubsection{Sequential IRS Deployment Method for a Given Subset of Candidate Sites}

In the following, we consider a subproblem of (P0) for any given subset of ${\cal I}_0$,  denoted as ${ \cal  I }_0^{*}$, with ${ \cal  I }_0^{*} \subseteq { \cal  I }_0$.
The method for selecting this high-quality subset of candidate sites will be discussed later. 
By replacing ${\cal I}_0$ in \eqref{Cons: Location2} with ${ \cal  I }_0^{*}$, the associated IRS deployment optimization subproblem can be formulated as 
\begin{subequations}
\begin{eqnarray} 
	\text{(P2):} \mathop{\text{Minimize}}\limits_{{\cal I}, {\cal T},{\mathcal S}} && c({\cal I}, {\cal T}) \nonumber\\
	\text{subject to}\;\;&&{\cal I} \subseteq { \cal  I }_0^{*}, \label{Cons: Location3}\\
	&& 	\text{Constraints~} \eqref{Cons:coveragerate}, \eqref{Cons:State}, \eqref{Cons:Tile}. \nonumber
\end{eqnarray}	 
  \end{subequations}
 
 Although (P2) can be reformulated as an ILP problem similar to (P1) and solved using the BB algorithm, we propose a more efficient sequential IRS deployment algorithm with a polynomial complexity to solve (P2). 
  Its basic idea is to sequentially select a candidate site from ${ \cal  I }_0^{*}$ for IRS deployment and  determine its tile number, height, and orientation in closed form, until the coverage rate constraint \eqref{Cons:coveragerate} is met.

To begin, we initialize ${\cal I} = \emptyset$, ${\cal S}= \emptyset$, and ${\cal T} = \emptyset$.
In each iteration,  a candidate IRS site is selected from ${ \cal  I }_0^{*}\setminus {\cal I}$ for deployment, and its associated tile number, height, and orientation are determined in closed form.
For example, consider that candidate site $i$, $i \in { \cal  I }_0^{*}\setminus {\cal I}$, is selected for IRS deployment in a specific iteration.
By substituting $\mathcal{I}\cup \{ i\},{\mathcal S} \cup \{\bm s_i\}$, and ${\mathcal T}\cup \{ T_i\}$ into \eqref{Eq:objective}, we have
\begin{eqnarray} \label{Eq:rate}
\!\!\!\!\!	&&\eta(\mathcal{I}\cup \{ i\},{\mathcal S} \cup \{\bm s_i\}, {\mathcal T}\cup \{ T_i\}) \nonumber \\
\!\!\!\!\!	&& = \frac{\sum_{n\in{\mathcal N}} u(T_i^2M^4\kappa_{0,i,n}(\bm s_i)  +	\bar{P}_{n}(\mathcal{I},{\mathcal S}, {\mathcal T})-P^{\min})}{N}.
\end{eqnarray}
From \eqref{Eq:rate}, it is evident that the value of $T_i^2M^4\kappa_{0,i,n}(\bm s_i)$ increases as $T_i$ increases since $M^4 \kappa_{0,i,n}(\bm s_i) $ is a non-negative coefficient for any $n$ and $i$.
Consequently, for any given  $\bm s_i$, $\eta(\mathcal{I}\cup \{ i\},{\mathcal S} \cup \{\bm s_i\}, {\mathcal T}\cup \{ T_i\})$ exhibits a piecewise-constant and non-decreasing behavior w.r.t. $T_i$ due to the property of the step function $u(\cdot)$. 
However, increasing the tile number $T_i$ also leads to 
a higher deployment cost $(\mathcal{I}\cup \{ i\},{\mathcal T}\cup \{ T_i\})$,  regardless of $\bm s_i$.
Thus,  the optimal height and orientation of  IRS $i$ (denoted as $\bm s_i^{\star}(T_i)$) should maximize the coverage rate $\eta(\mathcal{I}\cup \{ i\},{\mathcal S} \cup \{\bm s_i\}$, and ${\mathcal T}\cup \{ T_i\})$ for any given $T_i$, which is given by
\begin{eqnarray} \label{Eq:State1}
	\bm s_{i}^{\star}(T_i) \in \arg \max_{\bm s_i\in {\cal H}_i \times {\cal O}_i} \eta(\mathcal{I}\cup \{ i\},{\mathcal S} \cup \{\bm s_i\}, {\mathcal T}\cup \{ T_i\}).
\end{eqnarray}

With $\bm s_{i}^{\star}(T_i)$ for any $T_i$, the optimal tile number $T_i^{\star}$ should further maximize the coverage rate  $\eta(\mathcal{I}\cup \{ i\},{\mathcal S} \cup \{	\bm s_{i}^{\star}(T_i)\}, {\mathcal T}\cup \{ T_i\})$ while reducing the hardware cost (i.e., $c_{h,0} T_i$).
This  depends on whether the coverage rate constraint \eqref{Cons:coveragerate} in (P2) is satisfied by deploying this IRS.
The maximum achievable coverage rate is obtained by setting $T_i = T_0^{\max}$, and then denoted as  $\eta_i^{\max} \triangleq  \eta(\mathcal{I}\cup \{ i\},{\mathcal S} \cup \{\bm s_{i}^{\star}(T_0^{\max})\}, {\mathcal T}\cup \{ T_0^{\max}\})$.
By comparing $\eta_i^{\max}$ with the target  $\eta_0$, two outcomes are possible:
\begin{enumerate}[(i)]
\item If $\eta_i^{\max} \geq \eta_0$: In this case, deploying IRS $i$ can ensures that (P2) is feasible by optimizing $T_i$ and $\bm s_i$. 
Here,  we set the tile number $T_i$ as the minimum required value to satisfy $ \eta_{i}(T_i) \triangleq \eta(\mathcal{I}\cup \{ i\},{\mathcal S} \cup \{\bm s_{i}^{\star}(T_i)\}, {\mathcal T}\cup \{ T_i\}) \geq \eta_0$, thereby terminating the sequential IRS deployment procedure and avoiding additional site-use cost.

\item If $\eta_i^{\max} < \eta_0$: 
In this case, even deploying IRS $i$, additional candidate sites are required for IRS deployment to reach the coverage target. Here, the tile number $T_i$ is set to the minimum value that maximizes $ \eta_{i}(T_i)$, which may help reduce the number of additional IRSs needed to meet the coverage target.
\end{enumerate}

In summary, the tile number, height, and orientation of IRS deployed at candidate site $i$ are determined as 
\begingroup\makeatletter\def\f@size{9.2}\check@mathfonts\def\maketag@@@#1{\hbox{\m@th\normalsize\normalfont#1}}
\begin{eqnarray}
\!\!\!\!\!\!\!\!\!	&&T_{i}^{\star} \!=\! \nonumber \\
\!\!\!\!\!\!\!\!\!	&&\begin{cases}\!
		\min \{T_i \in \mathbb{Z}^{+}  \mid T_i\in [1,T_0^{\max }], \eta_{i}(T_i)\geq \eta_0 \}   &\!\!\!\!\! \text{if~}  \eta_i^{\max}\geq \eta_0,\\
		\!\min \left\{ T_i \in \mathbb{Z}^{+} \mid T_i \in \arg\max_{T_i} \eta_{i}(T_i) \right\}	&  \text{otherwise}, 
	\end{cases}\label{Eq:tileNumber}\\ 
\!\!\!\!\!\!\!\!\!	&& \bm s_{i}^{\star} = \bm s_{i}^{\star}(T_{i}^{\star}). \label{Eq:State}
\end{eqnarray} \normalsize
By applying \eqref{Eq:tileNumber} and \eqref{Eq:State}, the achievable coverage rate is maximized, while the hardware cost is minimized, thus improving the overall cost-performance tradeoff.

The above process is repeated iteratively, with the number of deployed IRSs (i.e., $|\mathcal{I}|$) increasing, and the achievable coverage rate non-decreasing  over the iterations. The selection of candidate sites for IRS deployment continues until the coverage rate constraint in (P2) is satisfied. However, it is important to note that the final performance critically depends on the order in which candidate sites for IRS deployment are selected.
We propose the following procedure to determine the order of selected candidate sites. In each iteration,  given the current sets ${\mathcal I}$, ${\mathcal S}$, and ${\mathcal T}$, we first determine $T_i^{\star}$ and $\bm s_{i}^{\star}$  for each $i  \in {\mathcal I}_0^{*} \setminus {\mathcal I}$ using \eqref{Eq:tileNumber} and \eqref{Eq:State}.
Then, for each $i  \in {\mathcal I}_0^{*} \setminus {\mathcal I}$, we have $\eta_i = \eta(\mathcal{I}\cup \{ i\},{\mathcal S} \cup \{\bm s_{i}^{\star}\}, {\mathcal T}\cup \{ T_i^{\star}\})$.
Next, we select one candidate site from ${\mathcal I}_0^{*} \setminus {\mathcal I}$ for IRS deployment based on the values of $T_i^{\star}$ and $\eta_i$.
We define ${\cal C} \triangleq  \{i| \eta_i \geq \eta_0, i\in  {\mathcal I}_0^{*} \setminus {\mathcal I}\}$.
If ${\cal C}\notin \emptyset$, we select one candidate site from ${\cal C}$, thereby terminating the procedure to avoid additional site-use cost.
All elements in ${\cal C}$ are sorted primarily by their $T_i^{\star}$'s values  in an ascending order  and secondarily by their $\eta_i$'s values in a descending order.
The site corresponding to the first element in this ordered list is selected for IRS deployment in this iteration, i.e., $i^\star = \arg\min_{i\in {\cal C}}\{T_i^{\star}, -\eta_i\}$.
If ${\cal C}= \emptyset$,  all elements in ${\mathcal I}_0^{*} \setminus {\mathcal I}$ are first sorted by their $\eta_i$ values in descending order and then by their $T_i^{\star}$'s values in an ascending order. The selected candidate site for IRS deployment is given by $i^\star = \arg\min_{i\in {\mathcal I}_0^{*} \setminus {\mathcal I}}\{-\eta_i, T_i^{\star}\}$.
The main procedures of the sequential IRS deployment method for solving (P2) are summarized in Algorithm \ref{Alg:1}.

\subsubsection{Successive Refinement Method for Solving (P0)}
Building upon the sequential IRS deployment method outlined earlier, we present the successive refinement method for solving (P0).
This approach iteratively solves a sequence of subproblems of (P0) via the sequential IRS deployment method by replacing ${\cal I}_0$  in \eqref{Cons: Location2} with successively refined and reduced subsets of  ${\cal I}_0$, until no further reduction in deployment cost is achievable.
Finally, we optimally solve a subproblem of (P0) with the most reduced subset of  ${\cal I}_0$  using the BB algorithm.

\textbf{Step 1: Finding a high-quality feasible solution of (P0) via sequential IRS deployment:} 
The first step involves identifying a subset of candidate IRS sites (i.e., ${ \cal  I }_0^{*}$) that effectively contribute to the objective function of (P0).
To achieve this, we set  $\chi_n = 1, \forall n$, while relaxing the binary variables $\xi^{i}_{tjk}$  in (P1)  to continuous variables.
The resulting linear programming problem is obtained as 
\begin{subequations}
\begin{eqnarray} 
	\text{(P2.1):} \mathop{\text{Minimize}}\limits_{\bm \Xi} && \sum_{i  \in {\cal I}_0}\sum_{t = 1}^{T_{0}^{\max}}\sum_{j = 1}^{|{\cal H}_i|}\sum_{k = 1}^{|{\cal O}_i|} (c_{P, 0} + c_{P}t) \xi^{i}_{tjk}  \nonumber\\
	\text{subject to}\;\;
	&& P^{\min} - \bar{P}_{n}(\bm \Xi)  \leq  0, \forall n \in {\cal N} \label{Cons:power2.1}\\ 
	&& \xi^{i}_{tjk}\in [0,1], \forall i \in { \cal  I }_0, \forall t,\forall j,\forall k.\\
	&& \text{Constraints}~\eqref{Cons:binaryvariable1}~,\eqref{Cons:coverIndicator}.
\end{eqnarray}	
\end{subequations}
Problem (P2.1) can be efficiently solved using the interior-point method  \cite{boyd2004convex} with a polynomial complexity.
If (P2.1) is feasible, the optimal solution is denoted as 
$\bm \Xi^{*} = \{(\xi^{i}_{tjk})^{*}\}$,  and we restrict the subset of ${ \cal  I }_0$ to
\begin{eqnarray} \label{eq:subset}
	{ \cal  I }_0^{*} = \{i \big| \sum_{t}\sum_{j}\sum_{k} (\xi^{i}_{tjk})^{*} \neq 0, i \in {\cal  I}_0 \}.
\end{eqnarray}	

Next, we substitute ${ \cal  I }_0^{*}$ into (P2) and solve it via Algorithm \ref{Alg:1} for obtaining its solution $\{{\cal I}^{*}, {\cal S}^{*}, {\cal T}^{*}\}$.
The subset of ${ \cal  I }_0$ is then updated as
\begin{eqnarray}
	{ \cal  I }_0^{*} \leftarrow {\cal I}^{*},
\end{eqnarray}	
which considerably reduces the problem size, making it computationally more manageable. In the following step, we successively refine the elements in ${ \cal  I }_0^{*}$ to further reduce the objective function of (P0).

\textbf{Step 2:  Successively refining the elements in ${ \cal  I }_0^{*}$:}
To further reduce the deployment cost, we propose to successively refine one element in ${\cal I}^{*}_0$ while fixing those of all other elements in it.
Specifically, let ${\cal I}^{*}_0 = \{i_1,\ldots, i_{|{\cal I}^{*}_0|}\}$.
When updating the $m$-th element in ${\cal I}_0^{*}$, the following optimization problem is solved,
\begin{subequations}
\begin{eqnarray} 
	\text{(P2.2):} \mathop{\text{Minimize}}\limits_{{\cal I}, {\cal T},{\mathcal S}, i'} && c({\cal I}, {\cal T}) \nonumber\\
	\text{subject to}\;\;&& i' \in { \cal  I }_0\setminus { \cal  I }_0^{*}, \\
	&& {\cal I} \subseteq \{i'\}\cup { \cal  I }_0^{*}\setminus \{i_m\}, \label{Cons: Location4}\\
	&& 	\text{Constraints~} \eqref{Cons:coveragerate}, \eqref{Cons:State}, \eqref{Cons:Tile}.
\end{eqnarray}	 
\end{subequations}
For any given $i'$, Algorithm \ref{Alg:1} is applied to solve (P2.2).
By enumerating all possible candidate sites in ${ \cal  I }_0\setminus { \cal  I }_0^{*}$,
the site yielding the smallest objective value of (P2.2) is selected, resulting in the new solution $ \{\tilde {\cal I}, \tilde{\cal S}, \tilde{\cal T}, \tilde{i}\}$.
If the new deployment cost is smaller than the incumbent one, i.e.,
$c(\tilde {\cal I}, \tilde{\cal T}) < c( {\cal I}^{*}, {\cal T}^{*})$,
we update   $ \{ {\cal I}^{*}, {\cal S}^{*}, {\cal T}^{*}\}$ with $ \{\tilde {\cal I}, \tilde{\cal S}, \tilde{\cal T}\}$ and update ${\cal I}^{*}_0$ by replacing $i_m$ with $\tilde{i}$.

With the updated $ \{ {\cal I}^{*}, {\cal S}^{*}, {\cal T}^{*}\}$ and ${\cal I}^{*}_0$,
we then proceed to update the next candidate site in ${\cal I}^{*}_0$ accordingly in a similar way.
Through this successive refinement, the total deployment cost $c( {\cal I}^{*}, {\cal T}^{*})$ will decrease or remain constant with each iteration. Once the successive refinement procedure is complete, the candidate subset ${ \cal I }_0^{*}$ is updated as
\begin{eqnarray}
	{ \cal  I }_0^{*} \leftarrow {\cal I}^{*}.
\end{eqnarray}	

\textbf{Step 3: Final refinement via solving an ILP:}
In the final step, we obtain the optimal solution to subproblem (P2) with the updated ${ \cal I }_0^{*}$ via the BB algorithm.
Similar to (P1), we define a set of binary variables $\hat{\bm \Xi} \triangleq \{\hat {\xi}^{i}_{tjk}\}, i \in {\cal I}_0^{*}$, $t\in \{1,\ldots,  T_{0}^{\max}\}$, $ j \in \{1,\ldots,  |{\cal H}_i|\}$, $k \in \{1,\ldots,  |{\cal O}_i|\}$ to specify possible deployment configurations at each candidate IRS site in ${ \cal I }_0^{*}$, and reformulate (P2) into the following ILP problem,
\begin{subequations}
\begin{eqnarray} 
	\text{(P2.3):}  \mathop{\text{Minimize}}\limits_{\hat{\bm \Xi}, \bm \chi} && \sum_{i  \in {\cal I}_0^{*}}\sum_{t = 1}^{T_{0}^{\max}}\sum_{j = 1}^{|{\cal H}_i|}\sum_{k = 1}^{|{\cal O}_i|} (c_{s, 0} + c_{h,0}t) \hat{\xi}^{i}_{tjk}  \nonumber\\
		\text{subject to}\;\;		
		&&	P^{\min}\chi_n- \bar{P}_{n}(\hat{\bm \Xi})  \leq 0, \forall n \in {\cal N},\\
		&& \hat{\xi}^{i}_{ijk}\in \{0,1\}, \forall i \in {\cal I}_0^{*}, \forall t, \forall j,  \forall k,\\
		&&	\chi_n \in \{0,1\}, \forall n \in {\cal N}, \\
		&& \text{Constraints} \eqref{Cons:CoverRate}, \eqref{Cons:coverIndicator},
	\end{eqnarray}	 
\end{subequations}
where $ \bar{P}_{n}(\hat{\bm \Xi}) =  \beta^2_{0,n} 
+ \sum_{i\in{{\cal I}_0^{*}}}\sum_{t = 1}^{T_{0}^{\max}}\sum_{j = 1}^{|{\cal H}_i|}\sum_{k = 1}^{|{\cal O}_i|} \alpha_{tjk}^{in} \hat{\xi}^{i}_{tjk}$.
Similar to (P1), we solve (P2.3) optimally by applying the BB algorithm. 

The main procedures of the successive refinement algorithm for solving (P0) are summarized in algorithm \ref{Alg:2} (Alg. \ref{Alg:2}). In addition to solving a sequence of sub-problems (P2.2) using the sequential IRS deployment method with a polynomial complexity, the main computational complexity of Alg. \ref{Alg:2} arises from solving problems (P2.1) and (P2.3).
Solving (P2.1) results in a computational complexity of ${O}(\sum_{i=1}^{|{\cal I}_0^*|}T_0^{\max} |{\cal H}_i| |{\cal O}_i|)$.
In contrast, solving (P2.3) incurs an exponential complexity on the order of $O(2^{(N + \sum_{i=1}^{|{\cal I}_0^*|}T_0^{\max} |{\cal H}_i| |{\cal O}_i|)})$ in the worst case.
However, it is important to note that the number of binary variables in (P2.3) is significantly smaller than that in (P1) due to the successive refinement of $|{\cal I}_0^*|$ in Steps 1 and 2.
Additionally, to solve (P2.3),  we use the solution $\{{\cal I}^{*},{\cal S}^{*}, {\cal T}^{*}\}$ obtained in step 2 as the initial point to expedite the convergence of the BB algorithm.
As a result, the computational complexity for solving (P2.3) via the BB algorithm remains manageable. Furthermore, since Alg. \ref{Alg:2} is executed offline, its overall computational complexity is practically affordable.
As will be demonstrated in Section \ref{Sec_Results}, Alg. \ref{Alg:2} achieves a comparable performance while significantly reducing the running time as compared to the optimal BB algorithm.

\section{Numerical Results}\label{Sec_Results}
Simulation results are presented in this section to evaluate the performance of our proposed multi-IRS deployment solutions by the optimal BB algorithm and  the suboptimal successive refinement algorithm (i.e., Alg. 2)
under real-world propagation environments, offering valuable insights into their practical application.
\begin{figure}[t]
	\centering
	\vspace{-1mm}
	\subfigure[Full region $\mathcal{D}_0$;]{\includegraphics[width=.4\textwidth]{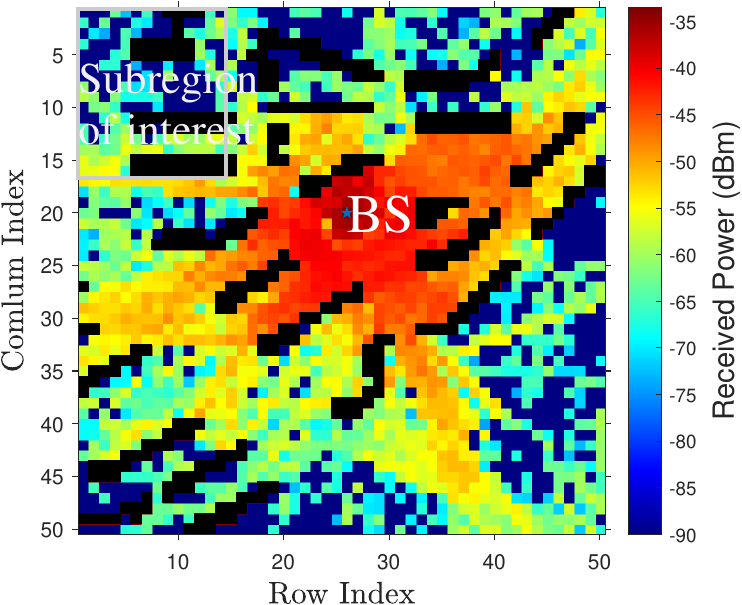}\label{Fig:withoutIRSFull}}
	\hspace{10mm}
	\vspace{-2mm}
	\subfigure[Sub-region of interest $\mathcal{D}$, $\mathcal{D}\subseteq \mathcal{D}_0$, $\frac{\sum_{n\in{\mathcal N}}u(\beta_{0,n}-P^{\min} )}{N} =  0.51$ with $P^{\min} = -68$ dBm.]{\includegraphics[width=.4\textwidth]{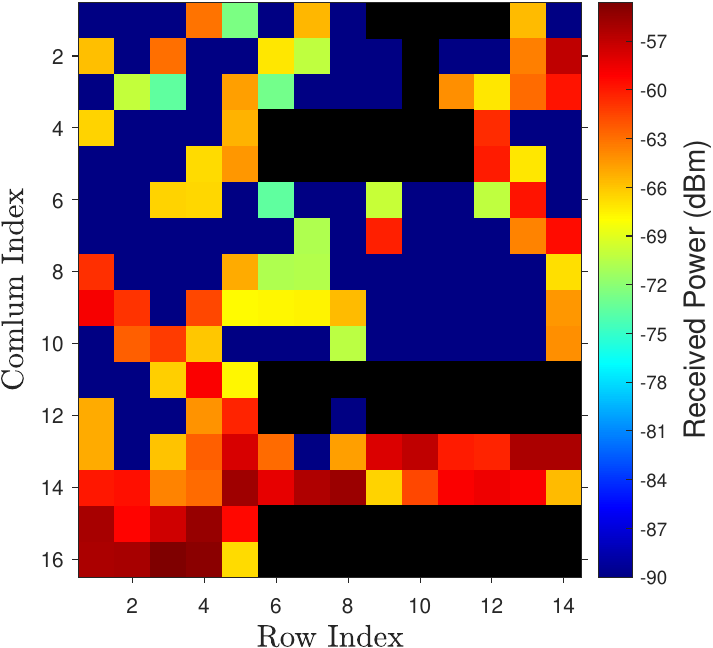}\label{Fig:withoutIRS}}
	\caption{Illustration of average channel power gain (i.e., $\beta_{0,n}$) maps from the BS without IRS in the outdoor propagation environment of Clementi, Singapore. The black areas represent building structures. } \label{FigWithoutmap}
	\vspace{-6mm}
\end{figure}

\begin{figure*}[t]
	\centering
	\subfigure[$\eta_0 = 1$, $T_0^{\max} = 25$, $c_{s,0} = 5$;]{\includegraphics[scale=0.5]{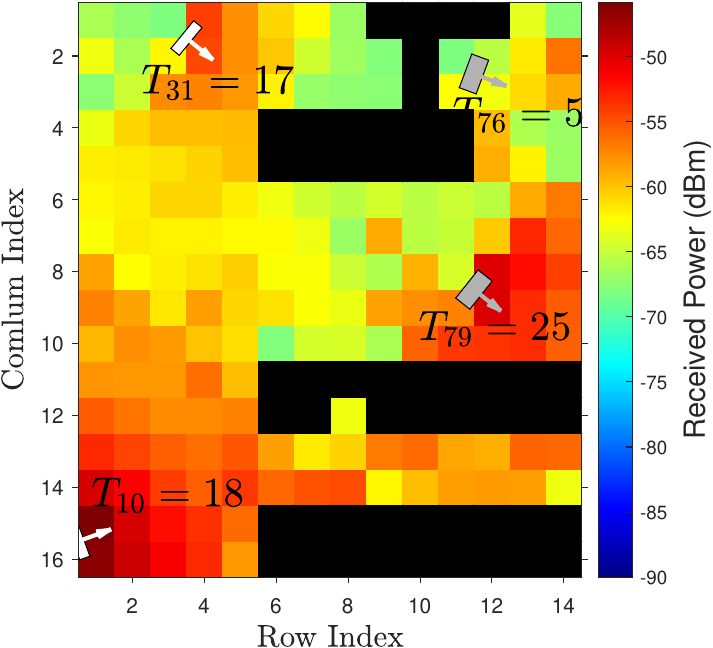}\label{Fig:MapEta1T25cp5}}
	\hspace{2mm}
	\subfigure[$\eta_0 = 0.9$, $T_0^{\max} = 25$, $c_{s,0} = 5$;]{\includegraphics[scale=0.54]{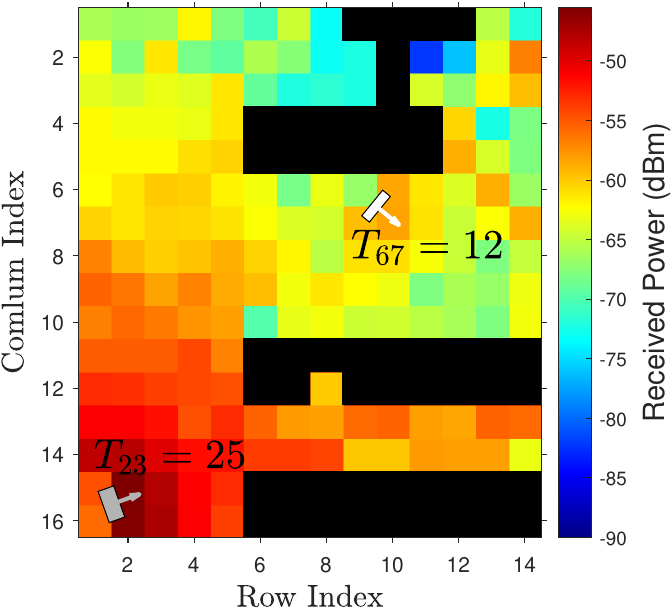}\label{Fig:Eta0.9T25cp5}}	
	\vspace{2mm}
	\subfigure[$\eta_0 = 1$, $T_0^{\max} = 16$, $c_{s,0} = 5$.]{\includegraphics[scale=0.5]{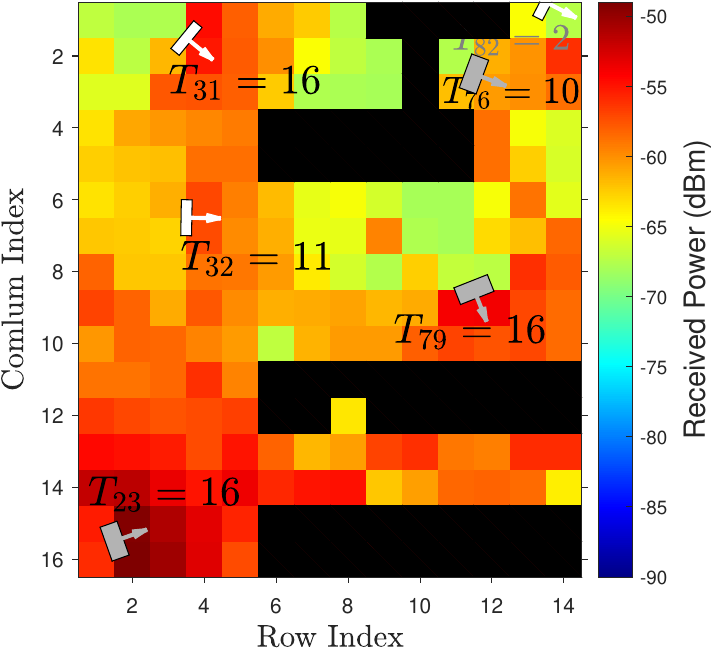}\label{Eta1T16cp5}}		
    \hspace{2mm}
	\subfigure[$\eta_0 = 1$, $T_0^{\max} = 25$, $c_{s,0} = 10$.]{\includegraphics[scale=0.5]{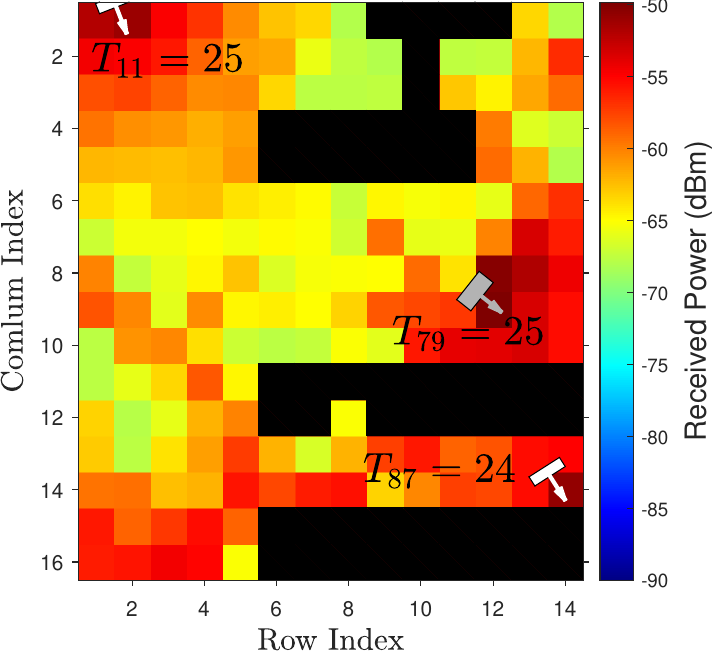}\label{Eta1T25cp10}}	
		\vspace{-4mm}
	\caption{Jointly optimized IRS deployment solutions under different values of $\eta_0$, $T_0^{\max}$, and $c_{s,0}$.}\label{Fig:Jointmap}
	\vspace{-6mm}
\end{figure*}

\subsection{Channel Power Measurements via Ray Tracing}\label{Sec:RayTracing}

To evaluate the performance of our proposed IRS deployment strategies, we employ ray tracing techniques to generate radio maps in real-world environments, measuring the average channel power gains of the BS-receiver links (i.e., $\beta_{0,n}$) and IRS-related links under different heights and orientations (i.e., $\|\bm{\sigma}_{0,i}(\bm{s}_i)\|^2$ and $\|\bm{\omega}_{i,n}(\bm{s}_i)\|^2$).

To simulate the signal propagation within the environment, we consider multiple transmission paths between any two points, focusing on specular reflections and diffracted paths while neglecting diffuse scattering due to the relatively large wavelength compared to the surface roughness of typical building materials. We use the shooting and bouncing ray (SBR) method to determine these paths, accounting for up to 8 reflections and 2 diffractions, which captures a wide range of possible interactions affecting signal strength. Additionally, we discard paths between any two points with power levels more than 6 dB below the maximum among all traced rays.

The simulation environment consists of a full region, $\mathcal{D}_0$, measuring 500 m $\times$ 500 m in Clementi, Singapore, with real-world building layouts sourced from OpenStreetMap. The entire region is discretized into 2500 grids with each size of 10 m $\times$ 10 m, where any grid center point with a height exceeding 8 m is considered as being occupied by buildings. The grid system begins at coordinate (1, 1) in the northwest corner, with positive coordinates extending eastward and southward, as illustrated in Fig. \ref{Fig:withoutIRSFull}.
In this setup, the BS is pre-deployed at grid (28, 21) and operates at a carrier frequency of 2.8 GHz, transmitting at a power of 30 dBm. 
A sub-region of interest, $\mathcal{D}$, is selected within $\mathcal{D}_0$, comprising grid rows indexed from 1 to 14 and columns indexed from 1 to 16, but excluding those grids occupied by buildings. 
This sub-region contains $|\mathcal{N}| = 171$ receiver grids, as depicted in Fig. \ref{Fig:withoutIRS}.

Given these simulation settings, we can first obtain the average channel power gains from the BS to receiver grids (i.e., $\beta_{0,n}$) using ray tracing in Matlab. The BS antenna height is set to 25 m, while receiver antennas in each grid are positioned at a height of 1.5 m, with isotropic antennas assumed for both transmitter and receiver.
However, a key challenge is that standard ray tracing platforms do not generally integrate IRS modeling. 
 To overcome this limitation and mimic IRS behaviors, we synthesize the average channel power gains of BS-to-IRS and IRS-to-grid links.
 First, we model each IRS as an isotropic transceiver antenna at each candidate IRS location to capture all rays originating from the BS and arriving at receiver grids. 
 Second, we enforce the half-space reflection constraint of  IRSs by only considering rays that arrive on the reflecting side for BS-to-IRS links, and rays that depart from the reflecting side for IRS-to-receiver links.
 Additionally, since isotropic antennas are used to simulate IRS elements, each path gain from one IRS to any receiver grid is scaled by a factor of 2 (or 3 dB) to account for the transmit half-space reflection gain inherent to IRSs. This approach enables us to realistically model IRS-related channel power gains and evaluate the impact of IRS deployment on network performance.

Fig. \ref{FigWithoutmap} illustrates the average channel power gain map from BS to each outdoor receiver grid without IRS, where black areas represent buildings for which no power measurement is performed.
In Fig. \ref{FigWithoutmap}, we set $\beta_{0,n} = -90$ dBm for any receiver grid $n$ where no direct link from the BS exists.
It is observed from Figs. \ref{Fig:withoutIRSFull} and \ref{Fig:withoutIRS} that the average channel powers of receiver grids in the sub-region of interest $\cal D$ are significantly lower than those closer to the BS satisfactorily. Most receiver grids within $\cal D$ experience severe signal degradation, rendering them as communication blind spots.
Specifically, when $P^{\min} = -68$ dBm, only 51$\%$ of receiver grids in $\cal D$ are covered by the BS, highlighting the need for enhanced coverage in this area. The deployment of additional IRSs within the sub-region is necessary to improve coverage and eliminate these blind spots.

In our simulation, we consider at most one candidate IRS site within each receiver grid. 
The possible heights for each IRS are set as ${\cal H}_i\triangleq\{10 , 15 \}$ in meter,  and the possible orientations are given by ${\cal O}_i \triangleq\{0 , \frac{\pi}{6}, \frac{-\pi}{6} \}$ in radian.
To identify candidate IRS sites, we propose the following approach. First, we select the location within each receiver grid $n$ that is farthest from the BS as a potential candidate site. Then, we assess whether this location can serve as a candidate IRS site by determining if it can establish a connection with the BS using any of the possible heights and orientations. If such a connection can be established, the location is confirmed as a candidate site. Otherwise, no candidate site is designated for that receiver grid. This process results in identifying $I_0 = 87$ candidate sites within the considered sub-region $\cal D$.

Other system parameters are set as follows, unless stated otherwise: the number of reflecting elements per tile is set to $M^2 = 256$, the minimum received power target is $P^{\min} = -68$ dBm, and the maximum number of tiles at each candidate location is $T_{0}^\text{max} = 25$. The hardware cost per element tile is set to $c_{h,0} = 1$, while the site-use cost per IRS is $\alpha_{s,0} = 5$.

We evaluate the performance of the proposed deployment strategies against two benchmarks:
\begin{itemize}
	\item {\it Fixed-State Deployment}: 
	This benchmark involves IRS deployment with minimal height and orientation directly facing the BS at all candidate locations. That is, the height and orientation are fixed at $h_{i}= 10$ m and $\theta_{i}= 0$ radian for all $i \in {\cal I}$ in (P0). The problem is reformulated as an ILP and solved using the BB algorithm, similar to (P1).
	
	\item {\it MaxTile Deployment}: This benchmark assumes each IRS is equipped with the maximum number of tiles. That is, the number of tiles is fixed at $T_{i}= T_0^{\max}$ for all $i \in {\cal I}$ in (P0), which is also solved via the BB algorithm after reformulated as an
	 ILP.

\end{itemize}

\subsection{Jointly Optimized IRS Deployment Solutions}

Fig. \ref{Fig:Jointmap} shows the optimal IRS deployment solutions for (P0) using the BB algorithm under the scenario with $P^{\min} = -68$ dBm, considering different values of $\eta_0$, $T_0^{\max}$, and $c_{s,0}$.
The candidate sites selected for deploying IRSs are indicated by white and gray rectangles, representing IRS heights of 10 m and 15 m, respectively. 
Each rectangle is accompanied by an arrow that indicates the IRS’s deployment orientation. Additionally, the number of tiles assigned to each IRS (denoted as $T_i$) is displayed in this figure, where the subscript $i$ in $T_i$ represents the one-dimensional index of the candidate site among a total of 87 candidate sites.

Specifically,  Figs. \ref{Fig:MapEta1T25cp5} and \ref{Fig:Eta0.9T25cp5} show the optimal IRS deployment solutions to (P0) under coverage rate targets $\eta_0 = 1$ and $\eta_0 = 0.9$, respectively, with $T_{0}^{\max} = 25$ and $c_{s,0} = 5$.
In Fig. \ref{Fig:MapEta1T25cp5}, the optimal solution deploys 4 IRSs with a total of 65 tiles to achieve full coverage, i.e., $\eta_0 = 1$. This result highlights that the proposed joint deployment strategy significantly enhances coverage performance compared to the scenario without IRS (i.e., $\eta^{\min} = 0.51$), and achieves the coverage goal with much fewer candidate sites selected as compared to deploying IRSs at all candidate sites.
Furthermore, Fig. \ref{Fig:Eta0.9T25cp5} shows that by reducing the coverage rate target from $\eta_0 = 1$ to $\eta_0 = 0.9$, the total number of deployed IRSs decreases from 4 to 2, and the total number of tiles decreases from 65 to 37.  
This demonstrates the performance-cost trade-off achievable by adjusting the coverage target.
In contrast to Fig. \ref{Fig:MapEta1T25cp5}, where $T_{0}^{\max} = 25$, Fig. \ref{Eta1T16cp5} shows the optimal deployment solution when the maximum number of tiles per IRS is reduced to $T_{0}^{\max} = 16$. 
In this case, the number of selected sites for IRS deployment increases from 4 to 6, indicating that increasing the number of tiles per IRS significantly reduces the number of selected sites required to achieve the same coverage target, thanks to the enhanced effective reflection gain by IRS.
Additionally, Fig. \ref{Eta1T25cp10} shows the optimal deployment solution for (P0) when the cost per site is increased to $c_{s,0} = 10$.
Compared to Fig. \ref{Fig:MapEta1T25cp5}, where $c_{s,0} = 5$, the number of sites with IRSs deployed decreases from 4 to 3, while the total number of tiles increases from 65 to 74. This result illustrates the trade-off between the site-use cost (i.e., reducing the number of selected sites) and hardware cost (i.e., the total number of tiles used). For any given $\eta_0$ and $T_0^{\max}$, there is a balance between minimizing the number of selected sites and reducing the total number of tiles, both of which contribute to decreasing the overall deployment cost.

\begin{figure}[t]
	\centering
	\includegraphics[width=.38\textwidth]{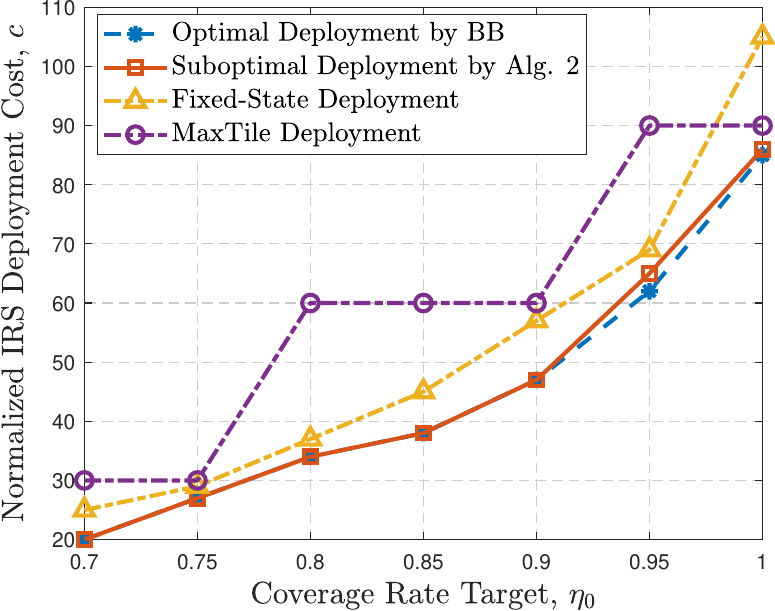}
	\vspace{-2mm}		
	\caption{Total deployment cost versus coverage rate under $c_{s,0} = 5$ and $T_{\max} = 25$. } \label{Fig:Cost-rate}
	\vspace{-2mm}
\end{figure}

\begin{table}[t]
	\centering
	\vspace{-2mm}
	\caption{Running time (in second) of algorithms for solving (P0).}\label{timetable(P0)}
	\begin{tabular}{|c|c|c|c|}
		\hline 
		&  \makecell[c]{$\eta_0 = 1$, \\ $T_0^{\max} = 25$} & \makecell[c]{$\eta_0 = 0.9$,\\ $T_0^{\max} = 25$} & \makecell[c]{$\eta_0 = 1$,\\ $T_0^{\max} = 16$} \\
		\hline
		BB Algorithm & $1.8\times10^4$ & $2.3\times10^5$ & $8.6\times10^4$\\
		\hline
		Alg. \ref{Alg:2} &  14.1   & 11.5  & 28.8 \\
		\hline
	\end{tabular}
	\vspace{-5mm}
\end{table}

\subsection{Performance Comparison}
Fig. \ref{Fig:Cost-rate} shows the total deployment cost by different deployment schemes versus the coverage rate target $\eta_0$. 
First, it is observed from Fig. \ref{Fig:Cost-rate} that the proposed suboptimal deployment by Alg. \ref{Alg:2} (i.e., successive refinement algorithm) achieves  a performance comparable to the optimal deployment by the BB algorithm.
This demonstrates the effectiveness of Alg. \ref{Alg:2} for solving (P0) in terms of minimizing the deployment cost.
To further evaluate the computational efficiency of Alg. \ref{Alg:2}, we also compare its running time with that of the BB algorithm in Table \ref{timetable(P0)}. 
The results reveal that Alg. \ref{Alg:2} achieves a significant reduction in running time, outperforming the BB algorithm by approximately 3 to 4 orders of magnitude. This marked reduction highlights the effectiveness and efficiency of Alg. \ref{Alg:2}, providing computationally efficient solutions to problem (P0) without sacrificing the coverage performance.
Additionally, it is observed from Fig. \ref{Fig:Cost-rate} that the proposed joint deployment strategy (i.e., the optimal or suboptimal deployment) incurs a substantially lower deployment cost than the fixed-state deployment, particularly when  $\eta_0 >0.8$. 
This is because the joint deployment method ensures that the effective reflection gains of  IRSs are maximized and the IRSs are positioned to provide optimal coverage, allowing the BS to establish stronger channel paths with the receiver grids via IRSs. 
This approach enhances coverage and reduces the number of sites and tiles required for IRS deployment, leading to significant savings in deployment cost.
Moreover, the MaxTile deployment strategy incurs the highest deployment cost among all strategies, with a particularly steep increase as the coverage rate target approaches 1. 
This shows the inefficiency of the MaxTile strategy,  due to over-provisioning and inefficient tile allocation.

\begin{figure}[t]
	\centering
	\includegraphics[width=.38\textwidth]{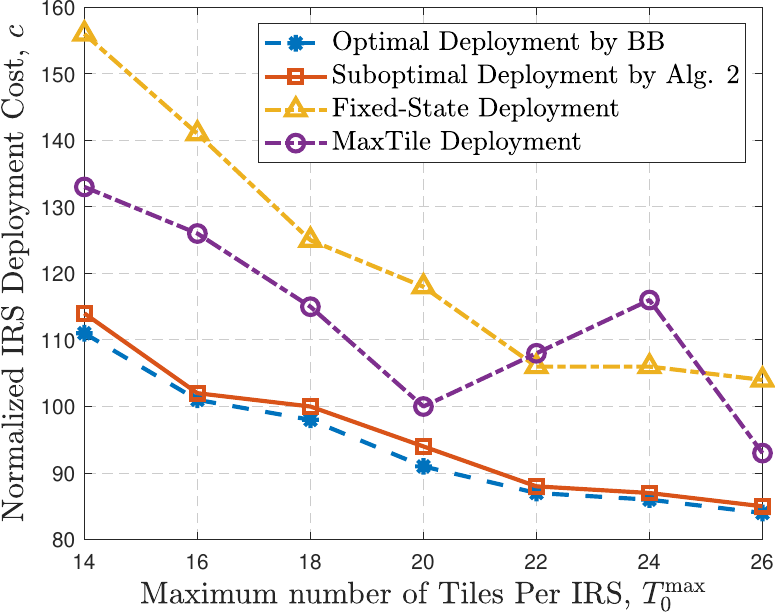}
	\vspace{-2mm}		
	\caption{Total deployment cost versus maximum number of tiles per IRS under $\eta_0 = 1$ and $c_{s,0} = 5$. } \label{Fig:Cost-tile}
	\vspace{-4mm}
\end{figure}

\begin{figure}[t]
	\centering
	\includegraphics[width=.37\textwidth]{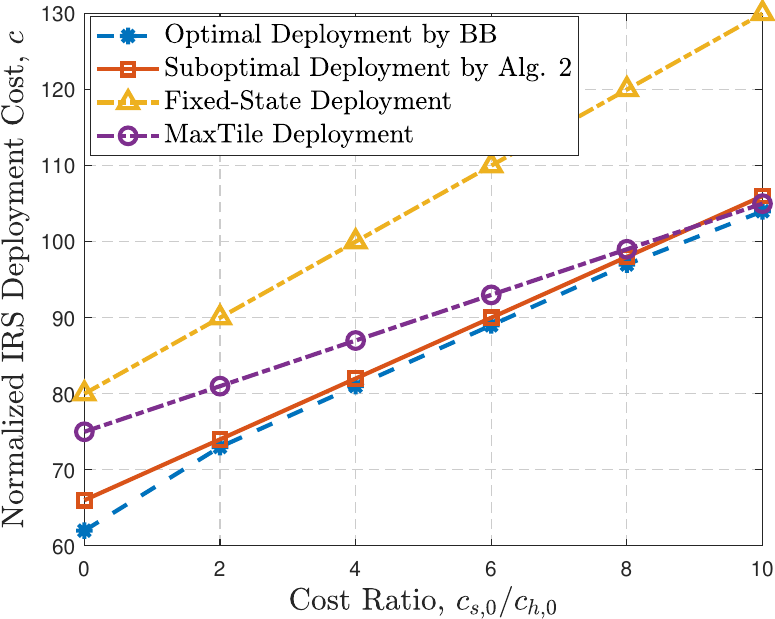}
	\vspace{-2mm}		
	\caption{Total deployment cost versus cost ratio under $\eta_0 = 1$ and $T_{\max} = 25$. } \label{Fig:Cost-ratio}
	\vspace{-6mm}
\end{figure}

Fig. \ref{Fig:Cost-tile} shows the deployment cost versus the maximum number of tiles per IRS, given $\eta_0 = 1$ and $c_{s,0} = 5$. 
The first observation is that the proposed  suboptimal deployment by Alg. \ref{Alg:2} achieves a near-optimal performance compared to the optimal deployment by the BB algorithm.
Second,  similar to both the proposed  optimal and suboptimal deployments, the deployment cost by the Fixed-State deployment decreases as $T_{0}^{\max}$ increases. 
This is because increasing the number of tiles per IRS improves its effective reflection gain (i.e., aperture and array gains), ensuring that more receiver grids are covered effectively. As a result, fewer IRS sites are needed, which reduces the overall deployment cost, given that the fixed cost per IRS site is much higher than the cost per tile.
Conversely, the MaxTile strategy does not show a consistent decrease in deployment cost as $T_{0}^{\max}$ increases. While there is an initial reduction in cost for $14 \leq T_{0}^{\max} \leq 20$, owing to the enhanced effective IRS reflection gain and fewer required deployment sites, this benefit diminishes as $T_{0}^{\max}$ increases beyond 20. 
At this point, the increased hardware cost associated with deploying more tiles outweigh the saving from reducing the number of IRS sites. Consequently, for $T_{0}^{\max} > 20$, the deployment cost begins to rise, illustrating the inefficiency of the MaxTile strategy at higher tile numbers.
In contrast, the proposed suboptimal deployment by Alg. 2 significantly outperforms both the Fixed-State and MaxTile strategies across the entire range of $T_{0}^{\max}$ by fully leveraging the degrees of freedom in optimizing deployment configurations. This flexibility allows the proposed approach to strike an optimal balance between site selection and tile usage, minimizing the overall deployment cost.

As previously discussed in Figs. \ref{Fig:MapEta1T25cp5} and \ref{Eta1T25cp10}, there exists a balance between site and tile numbers that minimizes the total deployment cost, depending on the ratio of $c_{s,0}$ to $c_{h,0}$. 
To further explore this trade-off, Fig. \ref{Fig:Cost-ratio} presents the total deployment cost as a function of the fixed cost for deploying one IRS ($c_{s,0}$), under $\eta_0 = 1$ and $T_{\max} = 25$.
Firstly, the proposed suboptimal deployment by Alg. 2 consistently shows performance close to the optimal deployment by the BB algorithm, particularly at higher cost ratios, underscoring its effectiveness. 
Secondly, the Fixed-State deployment strategy results in the highest deployment cost across all values of $c_{s,0}/c_{h,0}$, since inflexibility of IRSs' heights and orientations necessitates more sites for IRS deployment to form more paths for coverage rate improvement, thereby driving up the overall cost.
Additionally, it is observed from Fig. \ref{Fig:Cost-ratio} that the performance gap between the proposed suboptimal deployment by Alg. 2 and the MaxTile deployment narrows as the ratio of $c_{s,0}/c_{h,0}$ increases.
 In this context, the MaxTile deployment, which maximizes tile usage per site, becomes more competitive. Nonetheless, the proposed suboptimal deployment by Alg. 2 continues to outperform the MaxTile strategy by optimally balancing the number of sites and tiles across a wide range of cost ratios, ensuring more efficient and cost-effective IRS deployment.

\section{Conclusions}\label{Sec_Conclusion}

In this paper, we studied a multi-IRS deployment problem to enhance signal coverage in multi-path propagation environments by leveraging IRS reflections. 
We started by dividing the target area into multiple grids and derived the average channel power gains for the BS-IRS and IRS-receiver grid links, taking into account IRS deployment configurations, including position, size, height, and orientation.
We then modeled the deployment cost and characterized the coverage rate performance over all receiver grids based on large-scale channel knowledge only.
To optimally balance the performance-cost trade-off, we aimed to minimize the total deployment cost while ensuring a given coverage rate target by jointly optimizing the locations of IRSs and their deployment configurations (i.e., size, height, and orientation).
An optimal algorithm and a lower-complexity suboptimal algorithm were proposed to efficiently solve this combinatorial optimization problem. Numerical results demonstrated that the proposed suboptimal algorithm can achieve a near-optimal performance while significantly reducing the computational time compared to the proposed optimal algorithm. Additionally, the proposed joint IRS deployment strategy (with the optimal or suboptimal algorithm)
significantly outperforms other baseline deployment schemes, highlighting its effectiveness in achieving superior performance-cost trade-offs in realistic environments.

\bibliographystyle{IEEEtran}
\bibliography{Ref}

\end{document}